\documentclass[hidelinks,onefignum,onetabnum]{siamart250211}

\usepackage{lipsum}
\usepackage{amsfonts}
\usepackage{graphicx}
\usepackage{epstopdf}
\usepackage{algorithmic}
\ifpdf
\DeclareGraphicsExtensions{.eps,.pdf,.png,.jpg}
\else
\DeclareGraphicsExtensions{.eps}
\fi

\newsiamremark{remark}{Remark}
\newsiamremark{hypothesis}{Hypothesis}
\crefname{hypothesis}{Hypothesis}{Hypotheses}
\newsiamthm{claim}{Claim}
\usepackage{amsopn}

\ifpdf
\hypersetup{
pdftitle={A splitting variational quantum algorithm for the nonlinear Dirac equation},
pdfauthor={Q. Zuo, Y. He and X. Zhao}
}
\fi
\usepackage{amsmath,amssymb,amsfonts,mathrsfs,mathtools}
\usepackage{graphicx}
\usepackage{epsfig}
\DeclareGraphicsExtensions{.pdf,.png,.eps}
\usepackage{bm}
\usepackage{enumerate}
\usepackage{cases}
\usepackage{subeqnarray}
\usepackage{hyperref}
\usepackage{booktabs}
\usepackage{float}
\usepackage{xcolor}
\usepackage{tikz}
\usepackage{quantikz}
\allowdisplaybreaks

\newcommand{\fe}{\mathrm{e}}

\newcommand{\bR}{{\mathbb R}}

\newcommand{\bT}{{\mathbb T}}

\begin{document}

\headers{VQA for NLDE}{Q. Zuo, Y. He and X. Zhao}

\title{Numerical solution of the nonlinear Dirac equation by a splitting variational quantum algorithm \thanks{Submitted to the editors DATE.
\funding{Q. Zuo is supported by NSFC 12401488 and the Provincial Natural Science Research Project of Anhui Colleges under grant 2025AHGXZK30544. 
Y. He is supported by High-level Talent Research Foundation of Anhui Agricultural University under grant rc482504. 
X. Zhao is supported by National Key Research and Development Program of China, National MCF Energy R\&D Program (No. 2024YFE03240400) and NSFC 42450275, 12271413.
}}}

\author{Qian Zuo\thanks{School of Big Data and Statistics, Anhui University, Hefei 230601, China (\email{qianzuo@ahu.edu.cn}).}
\and Ying He\thanks{School of Information and Artificial Intelligence, Anhui Agricultural University, 230036 Hefei, China (\email{heymath@ahau.edu.cn}).}
\and Xiaofei Zhao\thanks{School of Mathematics and Statistics \& Computational Sciences Hubei Key Laboratory, Wuhan University, 430072, Wuhan, China
  (\email{matzhxf@whu.edu.cn}, \url{http://jszy.whu.edu.cn/zhaoxiaofei/en/index.htm}).}
}

\maketitle
\begin{abstract}
In this work, we propose an operator-splitting variational quantum algorithm, termed Dirac-sVQA, for simulating the nonlinear Dirac equation (NLDE). The main difficulty arises from the state-dependent nonlinear interaction, its time-discrete update depends explicitly on the intermediate spinor state and, in general, cannot be implemented as a fixed state-independent unitary circuit. To address this difficulty, we decompose the NLDE evolution into a structured linear Dirac substep and a nonlinear variational correction. The linear substep is implemented by a spinor-Fourier Dirac propagator on a joint position-spin register, preserving the spin-momentum coupling and mass-induced spin evolution of the Dirac operator. The nonlinear correction is reformulated as a measurement-based variational update through a small set of overlap, self-channel, and cross-channel observables. We provide the corresponding quantum circuits and derive measurement-aware resource and complexity estimates. Numerical experiments in several nonlinear regimes show that Dirac-sVQA accurately captures both the total density and the componentwise spinor dynamics, agrees well with classical Fourier pseudospectral splitting solutions, and exhibits stable error behavior over time. These results provide numerical evidence for the feasibility of operator-splitting variational quantum simulation for nonlinear relativistic wave equations.

\end{abstract}



\section{Introduction}\label{sec:introduction}
The Dirac equation was originally proposed by Paul Dirac in 1928 to describe spin-1/2 massive particles \cite{Dirac}. As the first mathematical framework to successfully unify quantum mechanics and special relativity, it has since found broad applications across multiple fields, such as quantum field theory \cite{lambda} and materials science \cite{Material}.
In this study, as a model problem, we consider the one-dimensional nonlinear Dirac equation (NLDE) \cite{Alvarez1983, semire1, Cuevas2014, Xu2013NLDE} on a torus $\bT=\bR/(2\pi)$ with periodic boundary conditions:
\begin{equation}\label{eq:nlde}
\begin{cases}
\displaystyle i \partial_t \Psi(t, x) =  -i\alpha\partial_x \Psi(t, x) +  \beta\Psi(t, x) + F\left(\Psi(t, x)\right), \, t > 0,\; x \in \bT,\\
\Psi(0,x)=\Psi^0(x), \, x\in\bT,
\end{cases}
\end{equation}
where $\Psi(t,x)=\left(\Psi_0(t,x),\Psi_1(t,x)\right)^T:[0,+\infty)\times\bT\to\mathbb{C}^2$ is the complex-valued vector wave function of the spinor field, and
\begin{equation*}
\alpha=
\begin{pmatrix}
0 & 1\\
1 & 0
\end{pmatrix},
\qquad
\beta=
\begin{pmatrix}
1 & 0\\
0 & -1
\end{pmatrix}
\end{equation*}
are the Pauli matrices. Equation \eqref{eq:nlde} constitutes a fundamental component of models in quantum electrodynamics, e.g.,  \cite{DKG, MDsplitting}. 
The nonlinearity $F(\Psi)$ in \eqref{eq:nlde} is usually taken as \cite{nonre3}
\begin{equation}
F(\Psi) = \lambda_1 (\Psi^{\ast} \beta \Psi)\beta\Psi + \lambda_2 |\Psi|^2 \Psi,
\end{equation}
with $|\Psi|^2 = \Psi^{\ast}\Psi$, where $\lambda_1, \lambda_2 \in \mathbb{R}$ are two given real constants, $\Psi^{\ast} = \overline{\Psi}^{\,T}$ is the complex conjugate transpose of $\Psi$.  
The chosen nonlinearity is physically motivated by distinct regimes. The case $\lambda_2 = 0$ and $\lambda_1 \neq 0$ corresponds to the Soler model from quantum field theory \cite{lambda}, whereas the case $\lambda_1 = 0$ and $\lambda_2 \neq 0$ describes a density-dependent nonlinearity widely used in the modeling of Bose–Einstein condensates \cite{nonlinear12}.
When both $\lambda_1 \neq 0$ and $\lambda_2 \neq 0$, the model describes a fully coupled nonlinear regime in which self-interaction and density-dependent interaction coexist, leading to an interplay between spin polarization and density modulation effects.

Classical numerical methods that have been extensively developed, e.g., \cite{nonre3, nldeua, MDsplitting, nldelow} for solving \eqref{eq:nlde}, require explicit storage and update over all spatial grid points, causing computational and memory costs to grow rapidly with problem dimension and resolution. 
By contrast, quantum algorithms encode large discrete states into quantum amplitudes, offering a more compact representation of high-dimensional solution spaces and a promising framework for PDEs \cite{Cerezo2021VQAReview, ingelmann2024two, jaksch2023variational, Montanaro2016, Tennie2025}.
Quantum algorithms for PDEs mainly fall into three categories: quantum linear-system approaches, which solve discretized PDEs in quantum-state form \cite{Cao2013Poisson, Childs2021HighPrecisionPDE}; direct quantum-evolution approaches, such as Schr{\"o}dingerisation, which reformulate linear differential equations into Schr{\"o}dinger-type systems \cite{Hu2024QuantumCircuitsPDE, Jin2024Schrodingerisation}; and variational or hybrid quantum-classical methods, which have been extended to nonlinear and multidimensional PDEs in the NISQ regime \cite{Alghassi2022FeynmanKac, Ali2023PoissonVQA, Lubasch2020Nonlinear, Sarma2024NonlinearMultiPDE, Yuan2019}. Nevertheless, most existing work still focuses on linear or weakly nonlinear equations, while strongly nonlinear, state-dependent, and non-unitary PDEs remain largely unexplored.

Quantum simulation of the Dirac equation has also attracted considerable attention. Existing studies include gate-based simulations \cite{Fillion2017DiracQC}, quantum-walk and cellular-automaton formulations \cite{Arrighi2013QWDirac, DAriano2013QCADirac, Miyashita2023DiracQW}, and experimental realizations on quantum platforms \cite{Gerritsma2010Dirac,Zhang2025DiracSC}. 
These works demonstrate that the linear Dirac equation is naturally compatible with unitary quantum dynamics. However, the nonlinear Dirac equation introduces a qualitatively different difficulty: its interaction term depends on the unknown quantum state itself and is generally nonunitary after time discretization. As a result, the nonlinear update cannot be implemented as a fixed state-independent quantum circuit, nor can it be evaluated pointwise from amplitude-encoded data as in
classical algorithms.

Recently, a hybrid variational quantum algorithm was proposed for the nonlinear Schr\"odinger equation \cite{kocher2025numerical}. By separating linear propagation from nonlinear correction, this strategy reduces the optimization burden and improves accuracy and long-time stability compared with a fully variational treatment of the entire PDE. Extending this idea from scalar Schr\"odinger-type equations to the NLDE is nontrivial. The NLDE involves a two-component spinor field, matrix-valued first-order transport, mass-induced spin evolution, spin-momentum coupling, and nonlinear interactions depending on both density and spin polarization. These features make the quantum formulation substantially more involved, requiring a coherent treatment of state encoding, structure-preserving linear propagation, nonlinear measurement design, and variational optimization.

In this work, we propose the Dirac-sVQA, an operator-splitting variational quantum algorithm tailored to the matrix-valued spinor structure of the NLDE. The method is built on a position-spin amplitude encoding and consists of two structure-aware components. The first is a spinor-Fourier Dirac propagator (SFDP), which implements the structured linear Dirac substep on a position-spin register through Fourier-basis propagation, signed Fourier momentum encoding, momentum-controlled spin rotations, and a local mass-induced spin rotation. The second is a Dirac quantum nonlinear processing unit (D-QNPU), which converts the state-dependent and generally nonunitary
nonlinear correction into measurable overlap, self-spin, and cross-spin channels. These measurement primitives yield an implementable variational cost function for preparing the spinor state at the next time step.

The main contributions of this work are as follows. First, we introduce a position-spin quantum encoding that represents both the spatial grid and the internal spin degrees of freedom of the discretized Dirac spinor. Second, we
construct the SFDP module, which preserves the characteristic spin-momentum coupling and mass-induced spin evolution of the linear Dirac operator at the circuit level. Third, we develop the D-QNPU framework for evaluating the
nonlinear Dirac correction through measurable Dirac-specific observables. Finally, we provide circuit-level resource estimates, measurement-aware complexity bounds, and numerical benchmarks against classical Fourier
pseudospectral splitting solvers in several nonlinear regimes. More broadly, these ingredients provide a transferable algorithmic framework for operator-splitting variational simulation of nonlinear relativistic spinor equations, including effective high-energy particle models with relativistic dispersion and nonlinear spinor interactions. This connection identifies NLDE-type models as a natural testbed for the future development of quantum algorithms for nonlinear relativistic dynamics.

The rest of the paper is organized as follows. In Section~\ref{sec:splitting}, we derive the operator-splitting formulation of the NLDE, including the Fourier-pseudospectral treatment of the linear Dirac
subflow and the explicit Euler discretizations of the nonlinear subflow. In Section~\ref{sec:Dirac-sVQA}, we develop the Dirac-sVQA framework, including the position-spin amplitude encoding, the SFDP linear substep, the D-QNPU self/cross-channel formulation, the measurable variational cost function, and the measurement complexity analysis. In Section~\ref{sec:circuit}, we present the explicit gate-level realization of the SFDP module and the D-QNPU
measurement circuits for the nonlinear observables. Numerical experiments in several nonlinear regimes are reported in Section~\ref{sec:numerics}, followed by conclusions and future directions in Section~\ref{sec:conclusion}.

\section{Operator splitting method}\label{sec:splitting}
In this section, we consider the operator splitting type methods \cite{MQ2002}, which have been considered for solving the NLDEs in the litersture \cite{nonre3, KSZ2021}, etc. Now, by combining this type of method with the VQA, we introduce a quantum algorithm to solve the NLDE \eqref{eq:nlde}.
The splitting method enables us to maintain a first-order time integration scheme for the nonlinear terms of the NLDE, while treating the linear part as an integrating factor combined with the Fourier transform, which makes the implementation of quantum circuits feasible.

We denote the torus $\bT=(a,b)$ in the NLDE \eqref{eq:nlde}. Let $L=b-a$ represent the computation interval and the spatial mesh size $h=\Delta x=\frac{L}{M}$ with $M$ being an even integer. The grid points are $x_j=a+jh, \ j=0,1,\dots,M-1$ and we use periodic boundary conditions such that $x_M \equiv x_0$. Such periodic boundary also serves as a valid domain truncation for the whole space problem, when the initially localized waves do not reached the boundary within the time of interests. For the spatial discretizations, owning to the periodic setup, it is convenient to apply the Fourier pseudospectral method \cite{spectral}.

It begins by splitting NLDE \eqref{eq:nlde} into two sub-flows:
\begin{equation}\label{eq:linear-flow}
\Phi_t^{\text{k}}: \quad i\partial_t \Psi
=-i\alpha\partial_x\Psi+\beta\Psi
=: \mathcal{L}\Psi, \quad t>0, \ x\in \bT
\end{equation}
and
\begin{equation}\label{eq:nonlinear-flow}
\Phi_t^{\text{p}}: \quad i\partial_t \Psi
=\lambda_1(\Psi^{\ast}\beta\Psi)\beta\Psi+\lambda_2|\Psi|^2\Psi
=: \mathcal{N}\Psi, \quad t>0, \ x\in \bT
\end{equation}
represent the linear and nonlinear parts, respectively.
For $\Phi_t^{\text{k}}$, we have the exact integration
\begin{equation}\label{exact linear}
    \Phi_t^{\text{k}}(\Psi(x,t)): \quad \Psi(x,t+\tau)=\fe^{-i\tau\mathcal{L}}\Psi(x,t),
\end{equation}
where $\tau=\Delta t$ denote the time step. 
For $\Phi_t^{\text{p}}$, traditional numerical methods typically perform explicit, implicit or exact integration updates directly at each spatial grid point, and then by different compositions of these two parts yield different splitting methods \cite{HWYY2024, MDsplitting}. 
From the viewpoint of quantum implementation, the splitting separates the NLDE evolution into two algorithmically distinct tasks. The linear Dirac subflow admits a unitary Fourier-space realization that preserves the spin-momentum coupling and mass-induced spin rotation, leading to the spinor-Fourier Dirac propagator used in Dirac-sVQA. In contrast, the nonlinear subflow is state-dependent and, after time discretization, generally does not define a fixed unitary circuit; nor can its local amplitudes be accessed pointwise from an amplitude-encoded state as in a classical solver. We therefore reformulate the nonlinear correction in terms of measurement-accessible quantities and incorporate it through a variational update.

For the linear part $\Phi_t^{\text{k}}$, we consider directly using the exact integration \eqref{exact linear} for the computation. For the nonlinear part $\Phi_t^{\text{p}}$, many numerical discretizations can be further applied. 
Here, in order to rewrite it as a measurable expectation value-driven variational optimization problem, we consider using the first order Euler step to compute the explicit step of the nonlinear operator:
\begin{equation}\label{euler nonlinear}
    \Phi_t^{\text{p}}(\Psi(x,t)): \quad \Psi(x,t+\tau)=\Psi(x,t)-i\tau \mathcal{N}\Psi(x,t) + \mathcal{O}(\tau^2).
\end{equation}
Then by different compositions of $\Phi_t^{\text{k}}$ and $\Phi_t^{\text{p}}$, we obtain the different
splitting methods \cite{splitting1, MQ2002}.
The classical first order composition gives the first order splitting scheme for the NLDE \eqref{eq:nlde}:
\begin{equation}\label{lie splitting}
    \Psi(x,t+\tau)=\Phi_{\tau}^{\text{p}} \circ \Phi_{\tau}^{\text{k}}(\Psi(x,t)),\quad t=0,\tau,2\tau,\cdots.
\end{equation}

For the function $\Psi(x,t)$, $x\in\bT$, we compute its function value $\Psi_{\mathrm{num}}(x_j,t)\approx\Psi(x_j,t)$ on the spatial grids for $j=0,1,\cdots,M-1$ by the discrete Fourier transform and the wavenumbers of the Fourier modes are given by $\mu_\ell=\frac{2\pi\ell}{L},\ell=-\frac{M}{2},\cdots,\frac{M}{2}-1$. The time interval from $t$ to $t + \tau$ is divided into two sub-steps and is expressed in the following form in classical notation.
\begin{itemize}
    \item The first implicit substep is treated in Fourier pseudospectral form as
\begin{equation}\label{eq:linear-step}
\widetilde{\Psi}_{\mathrm{num}}(x_j,t)
=
\fe^{-i\tau \beta }\mathcal{F}^{-1}
\left(
\fe^{-i\alpha\mu_\ell \tau }
\mathcal{F}\bigl(\Psi(x_j,t)\bigr)
\right),
\end{equation}
where $\mathcal{F}$ and $\mathcal{F}^{-1}$ denote the discrete Fourier transform and its inverse, respectively.
\item The second explicit (Euler) substep follows to
\begin{align}\label{eq:nonlinear-step}
\Psi_{\mathrm{num}}(x_j,t+\tau)
=&
\widetilde{\Psi}_{\mathrm{num}}(x_j,t)-i\tau \Big[
\lambda_1\bigl(\widetilde{\Psi}_{\mathrm{num}}^*(x_j,t)\beta\widetilde{\Psi}_{\mathrm{num}}(x_j,t)\bigr)
\beta\widetilde{\Psi}_{\mathrm{num}}(x_j,t)
\nonumber\\
&
+\lambda_2|\widetilde{\Psi}_{\mathrm{num}}(x_j,t)|^2\widetilde{\Psi}_{\mathrm{num}}(x_j,t)
\Big].
\end{align}
\end{itemize}

The splitting formulation above also determines the architecture of the quantum algorithm developed in the next section. The linear Dirac substep possesses a unitary Fourier-space structure: the spatial derivative is
diagonal in the Fourier basis, the Dirac kinetic coupling induces a momentum-dependent rotation on the spin degree of freedom, and the mass term acts as a local spin rotation. This motivates a structure-preserving spinor-Fourier Dirac propagator on a joint position-spin register. In contrast, the nonlinear Euler substep is local in physical space but depends explicitly on the intermediate spinor state and is generally not a fixed unitary map. We therefore design a measurement-based variational treatment for the nonlinear correction, in which the required nonlinear information is compressed into a small set of measurable overlap, self-channel, and cross-channel observables. Thus, the operator-splitting scheme is not merely a
classical time discretization, but also the organizing principle that connects the NLDE integrator with the proposed hybrid quantum-classical Dirac-sVQA framework.

\section{The Dirac-sVQA method}\label{sec:Dirac-sVQA}

In this section, we develop the quantum realization of the split-step
discretization \eqref{eq:linear-step}-\eqref{eq:nonlinear-step}. Each time step is decomposed into a structured linear Dirac substep and a state-dependent nonlinear Euler correction. The two-component spinor is encoded on a joint position-spin register, on which the linear substep is implemented as a unitary circuit. 
In contrast, the nonlinear correction depends explicitly on the current intermediate state and is generally not a fixed unitary map, so it is treated through a measurement-based variational update. This hybrid construction defines the proposed Dirac-sVQA method, whose computational workflow is organized around two coupled modules. The SFDP module implements the structure-preserving linear Dirac substep by exploiting its unitary Fourier-space spin-momentum structure, whereas the D-QNPU module incorporates the state-dependent nonlinear correction through measurement-accessible self- and cross-channel observables within a variational update.

\subsection{Spinor amplitude encoding}\label{subsec:amp-encoding}

We first specify how the discretized Dirac spinor is represented on the quantum register and how the nonlinear Euler update is rewritten in amplitude-encoded form. As in the Fourier pseudospectral splitting scheme introduced in Section~\ref{sec:splitting}, we discretize the spatial interval into $M=2^n$ grid points, where $n$ is the number of position qubits, and evolve in time with step size $\tau$.

At a fixed time $t$, let
\begin{equation*}
\Psi_{\mathrm{num}}(x_j,t)
=
\bigl(\Psi_{\mathrm{num},0}(x_j,t),\Psi_{\mathrm{num},1}(x_j,t)\bigr)^T,
\qquad j=0,\dots,M-1,
\end{equation*}
denote the spatially discretized numerical Dirac spinor field. In the quantum algorithm, this discrete spinor is encoded into a normalized state on a composite position-spin register: the position degree of freedom is represented by $n$ qubits, while one additional qubit stores the two spinor components. Thus, after normalization, the discrete field $\Psi_{\mathrm{num}}(t)$ is represented by an $(n+1)$-qubit state $|\bm{\psi}_{\mathrm{num}}(t)\rangle$. The amplitude encoding introduced below makes this correspondence explicit in a form suitable for variational state preparation and quantum measurement.

To establish a uniform continuous-to-discrete normalization, assume that
\begin{equation*}
\int_{-\pi}^{\pi} |\Psi(x,t)|^2\,dx = \gamma,
\end{equation*}
where $\gamma$ denotes the conserved mass of the continuous solution. 
In the discrete implementation, $\gamma$ is understood as the corresponding
discrete mass
\begin{equation*}
\gamma
=
h\sum_{j=0}^{M-1}
|\Psi_{\mathrm{num}}(x_j,t)|^2
=
h\sum_{j=0}^{M-1}\sum_{\sigma=0}^{1}
|\Psi_{\mathrm{num},\sigma}(x_j,t)|^2,
\end{equation*}
so that the encoded quantum state is normalized exactly.
In the numerical experiments below, the initial data are normalized so that $\gamma=1$.
Throughout the quantum circuit construction, we use the position-first
convention
$|j,\sigma\rangle := |j\rangle\otimes|\sigma\rangle$,
where $j\in\{0,\ldots,M-1\}$ labels the spatial grid point and
$\sigma\in\{0,1\}$ labels the spin component. Thus the first $n$
qubits form the position register and the last qubit represents the
Dirac spin degree of freedom.

We define the discrete quantum amplitudes by
\begin{equation}\label{eq:dirac-amp-encoding}
\psi_{\mathrm{num},j,\sigma}(t)
=
\sqrt{\frac{h}{\gamma}}\,
\Psi_{\mathrm{num},\sigma}(x_j,t),
\qquad
\sigma\in\{0,1\},\quad j=0,\dots,M-1,
\end{equation}
and obtain the normalized quantum state
\begin{equation}\label{eq:dirac-quantum-state}
|\bm{\psi}_{\mathrm{num}}(t)\rangle
=
\sum_{j=0}^{M-1}\sum_{\sigma=0}^1
\psi_{\mathrm{num},j,\sigma}(t)\,|j,\sigma\rangle,
\, {\rm and} \, 
\langle \bm{\psi}_{\mathrm{num}}(t)| \bm{\psi}_{\mathrm{num}}(t)\rangle=1.
\end{equation}
This establishes the correspondence between the spatially discretized Dirac spinor and the quantum state prepared on the $(n+1)$-qubit register.

For notational simplicity, we suppress the subscript ``num'' in the quantum amplitudes whenever no confusion can arise, and write
\begin{equation}\label{eq:dirac-quantum-state-short}
|\bm{\psi}(t)\rangle
=
\sum_{j=0}^{M-1}\sum_{\sigma=0}^1
\psi_{j,\sigma}(t)\,|j,\sigma\rangle.
\end{equation}

After the linear Fourier-pseudospectral substep, let
$\widetilde{\Psi}_{\mathrm{num}}(x_j,t)$ denote the intermediate spinor
field. The nonlinear Euler update is local in position space and can be
written as
\begin{equation}\label{eq:dirac-Fj-classical}
\Psi_{\mathrm{num}}(x_j,t+\tau)
=
Q_j\,\widetilde{\Psi}_{\mathrm{num}}(x_j,t),
\end{equation}
where $Q_j
=
\mathbb I_s
-
i\tau
\left[
\lambda_1
\bigl(
\widetilde{\Psi}_{\mathrm{num}}^*(x_j,t)\beta\,
\widetilde{\Psi}_{\mathrm{num}}(x_j,t)
\bigr)\beta
+
\lambda_2
|\widetilde{\Psi}_{\mathrm{num}}(x_j,t)|^2
\mathbb I_s
\right]$.
Here $\mathbb I_s$ denotes the identity on the two-dimensional spin
space. Thus, the nonlinear step is diagonal in the position basis and acts
through a local $2\times2$ spin block at each grid point.

Using the amplitude encoding \eqref{eq:dirac-amp-encoding}, we rewrite
\begin{equation}\label{eq:dirac-tilde-spinor-encoding}
\widetilde{\Psi}_{\mathrm{num}}(x_j,t)
=
\sqrt{\frac{\gamma}{h}}\,
\widetilde{\bm{\psi}}_j(t),
\qquad
\widetilde{\bm{\psi}}_j(t)
=
\begin{pmatrix}
\widetilde{\psi}_{j,0}(t)\\[1mm]
\widetilde{\psi}_{j,1}(t)
\end{pmatrix},
\end{equation}
so that
\begin{equation}\label{eq:dirac-beta-term-encoding}
\widetilde{\Psi}_{\mathrm{num}}^*(x_j,t)\beta\,
\widetilde{\Psi}_{\mathrm{num}}(x_j,t)
=
\frac{\gamma}{h}\,
\widetilde{\bm{\psi}}_j^{\ast}(t)\beta\,
\widetilde{\bm{\psi}}_j(t),
\end{equation}
and
\begin{equation}\label{eq:dirac-density-term-encoding}
|\widetilde{\Psi}_{\mathrm{num}}(x_j,t)|^2
=
\frac{\gamma}{h}\,
\widetilde{\bm{\psi}}_j^{\ast}(t)\widetilde{\bm{\psi}}_j(t).
\end{equation}
Substituting \eqref{eq:dirac-beta-term-encoding} and
\eqref{eq:dirac-density-term-encoding} into
\eqref{eq:dirac-Fj-classical}, we obtain the amplitude-encoded local block
\begin{equation}\label{eq:dirac-Fj-quantum}
Q_j
=
\mathbb I_s
-
i\tau \,\frac{\gamma}{h}
\left[
\lambda_1
\bigl(
\widetilde{\bm{\psi}}_j^{\ast}(t)\beta\,
\widetilde{\bm{\psi}}_j(t)
\bigr)\beta
+
\lambda_2
\bigl(
\widetilde{\bm{\psi}}_j^{\ast}(t)
\widetilde{\bm{\psi}}_j(t)
\bigr)\mathbb I_s
\right].
\end{equation}
The corresponding global nonlinear map is therefore
$Q[\widetilde{\bm{\psi}}(t)]
=
\sum_{j=0}^{M-1}
|j\rangle\langle j|
\otimes
Q_j$.
Equivalently, this state-dependent map can be written as
\begin{equation}\label{eq:dirac-Q}
Q[\widetilde{\bm{\psi}}(t)]
=
\mathbb I_{xs}
-
i\tau\,G[\widetilde{\bm{\psi}}(t)],
\end{equation}
where
$\mathbb I_{xs}=\mathbb I_x\otimes \mathbb I_s$, and $\mathbb I_x=\sum_{j=0}^{M-1}|j\rangle\langle j|$.
The nonlinear operator is given by
\begin{equation}\label{eq:dirac-G}
G[\widetilde{\bm{\psi}}(t)]
=
\frac{\gamma}{h}
\sum_{j=0}^{M-1}
|j\rangle\langle j|
\otimes
\left[
\lambda_1
\bigl(
\widetilde{\bm{\psi}}_j^{\ast}(t)\beta\,
\widetilde{\bm{\psi}}_j(t)
\bigr)\beta
+
\lambda_2
\bigl(
\widetilde{\bm{\psi}}_j^{\ast}(t)
\widetilde{\bm{\psi}}_j(t)
\bigr)\mathbb I_s
\right].
\end{equation}
For a fixed intermediate state $\widetilde{\bm{\psi}}(t)$,
$Q[\widetilde{\bm{\psi}}(t)]$ is a linear matrix on the position-spin
Hilbert space. As an update rule, however, it is state dependent because its
coefficients are determined by the current intermediate amplitudes. Therefore,
it cannot be implemented as a fixed state-independent unitary circuit. Since
each local spin block in $G[\widetilde{\bm{\psi}}(t)]$ is Hermitian,
$G[\widetilde{\bm{\psi}}(t)]$ is Hermitian for each fixed
$\widetilde{\bm{\psi}}(t)$, whereas the Euler map
\eqref{eq:dirac-Q} is not exactly unitary. This representation is the starting
point for the measurable variational formulation developed below.

\subsection{SFDP: Structured linear Dirac substep on a position-spin register}
\label{subsec:SFDP}

We next consider the linear part of the split-step evolution. Although this
stage is followed by a variational nonlinear correction, it is not merely a
preprocessing step. Rather, it provides the relativistic backbone of the
algorithm. In contrast to Schr\"odinger-type propagation, which is typically
generated by a scalar second-order dispersive operator, the linear Dirac
substep is governed by a matrix-valued first-order operator that couples the
momentum and spin degrees of freedom and contains an explicit mass term.
Preserving this structure at the circuit level is essential for a quantum
solver designed for relativistic spinor dynamics.

Under the amplitude encoding \eqref{eq:dirac-quantum-state-short}, the Dirac
spinor is stored on a joint $(n+1)$-qubit position-spin register, consisting
of $n$ position qubits and one spin qubit. In accordance with the
position-first convention introduced in Section~\ref{subsec:amp-encoding}, the
linear intermediate state is written as
\begin{equation}
|\widetilde{\bm{\psi}}(t)\rangle
=
U_{\mathrm{SFDP}}(\tau)|\bm{\psi}(t)\rangle,
\end{equation}
where $U_{\mathrm{SFDP}}(\tau)$ denotes the structured unitary splitting block
used for the linear Dirac substep.

The key observation is that the spatial derivative becomes diagonal after a
quantum Fourier transform on the position register. Hence, the kinetic part of
the Dirac generator can be implemented in momentum space by momentum-dependent
rotations on the spin qubit, while the mass contribution remains local in the
spin sector. More precisely, the kinetic block is given by
\begin{equation*}
U_{\mathrm{kin}}(\tau)
=
(\mathcal{F}_x^{-1}\otimes \mathbb{I}_s)
\left[
\sum_{\ell}
|\ell\rangle\langle \ell|
\otimes
\exp(-i\tau\mu_\ell\alpha)
\right]
(\mathcal{F}_x\otimes \mathbb{I}_s),
\end{equation*}
where $\mathcal{F}_x$ is the quantum Fourier transform on the position
register, $\mu_\ell$ is the discrete Fourier momentum associated with mode
$\ell$, and $\alpha$ acts on the spin qubit. The mass block is
\begin{equation*}
U_m(\tau)
=
\mathbb{I}_x\otimes \exp(-i\tau\beta),
\end{equation*}
where $\beta$ is the spin-space Dirac matrix and $\mathbb{I}_x$ denotes
the identity on the position register. The SFDP module used in this work is
therefore
\begin{equation*}
U_{\mathrm{SFDP}}(\tau)
=
U_m(\tau)U_{\mathrm{kin}}(\tau).
\end{equation*}

This yields a gate-compatible unitary realization of the linear Dirac substep,
preserving its first-order transport, spin-momentum coupling, and
mass-induced spin rotation. Given a variational state-preparation circuit
$U(\bm{\theta}_t)$ for the state at time $t$, the corresponding
intermediate-state map is
\begin{equation*}
|\widetilde{\bm{\psi}}(t)\rangle
=
U_{\mathrm{SFDP}}(\tau)U(\bm{\theta}_t)
|0\rangle^{\otimes(n+1)}
=
\widetilde{U}(\bm{\theta}_t)|0\rangle^{\otimes(n+1)},
\end{equation*}
with $\widetilde{U}(\bm{\theta}_t) := U_{\mathrm{SFDP}}(\tau)U(\bm{\theta}_t)$.
The explicit gate decomposition of the SFDP block is presented in
Section~\ref{sec:circuit}.

\subsection{D-QNPU: Variational nonlinear update via measurable self/cross channels}\label{subsec:D-QNPU}
The map $Q[\widetilde{\bm{\psi}}(t)]$ depends explicitly on the intermediate
state. For a fixed intermediate state it is a linear matrix, but as an update
rule it is state dependent and therefore does not define a fixed
state-independent unitary circuit. Moreover, the Euler approximation
$Q[\widetilde{\bm{\psi}}(t)]$ is not exactly unitary. A direct gate-level implementation of this map would therefore require state-dependent amplitude information at every spatial grid point, which is incompatible with an efficient fixed-circuit realization if approached naively.
To overcome this difficulty, we introduce a \emph{Dirac quantum nonlinear processing unit} (D-QNPU). The purpose of the D-QNPU is not to realize the nonlinear map $Q[\widetilde{\bm{\psi}}(t)]$ itself as a standalone unitary block, but to extract, through quantum measurements, the nonlinear information required for a variational update. 

Let $a_j=|\widetilde{\psi}_{j,0}(t)|^2$, and $b_j=|\widetilde{\psi}_{j,1}(t)|^2$. We introduce the two state-dependent diagonal operators
\begin{equation*}
D_{\mathrm{self}}
=
\sum_{j=0}^{M-1}
|j\rangle\langle j|
\otimes
\begin{pmatrix}
a_j & 0\\
0 & b_j
\end{pmatrix},
\end{equation*}
and
\begin{equation*}
D_{\mathrm{cross}}
=
\sum_{j=0}^{M-1}
|j\rangle\langle j|
\otimes
\begin{pmatrix}
b_j & 0\\
0 & a_j
\end{pmatrix}.
\end{equation*}
Equivalently, under the position-first convention,
\begin{equation*}
D_{\mathrm{cross}}=X_S D_{\mathrm{self}}X_S,
\qquad
X_S= \mathbb{I}_x^{\otimes n}\otimes X_s.
\end{equation*}
For a trial updated state $|\bm{\psi}(t+\tau)\rangle$, define
\begin{equation*}
E_{\mathrm{self}}
=
\mathrm{Im}
\langle \bm{\psi}(t+\tau)|
D_{\mathrm{self}}
|\widetilde{\bm{\psi}}(t)\rangle,
\end{equation*}
and
\begin{equation*}
E_{\mathrm{cross}}
=
\mathrm{Im}
\langle \bm{\psi}(t+\tau)|
D_{\mathrm{cross}}
|\widetilde{\bm{\psi}}(t)\rangle.
\end{equation*}
Thus, the nonlinear contribution admits the self-cross decomposition
\begin{equation}\label{eq:dirac-nonlinear-decomp}
\mathrm{Im}\,
\langle \bm{\psi}(t+\tau)|
G[\widetilde{\bm{\psi}}(t)]
|\widetilde{\bm{\psi}}(t)\rangle
=
\frac{\gamma}{h}
\Big[
(\lambda_2+\lambda_1)E_{\mathrm{self}}
+
(\lambda_2-\lambda_1)E_{\mathrm{cross}}
\Big].
\end{equation}
Here, $E_{\mathrm{self}}$ and $E_{\mathrm{cross}}$ represent the intra- and inter-component nonlinear contributions, respectively. In the self channel, each spinor component is weighted by its own intermediate-state density, whereas in the cross channel it is weighted by the density of the opposite component. This self-cross decomposition exploits the intrinsic two-component spinor structure of the NLDE and is therefore specific to the Dirac
formulation.

Both observables can be estimated by dedicated measurement circuits acting on the same position-spin register, thereby avoiding full state tomography and pointwise classical reconstruction of the nonlinear field variables. Consequently, the nonlinear information required at each time step is compressed into a small set of measurable global observables. The D-QNPU provides these observables for evaluating the variational cost, while the updated spinor state is obtained by optimizing a parameterized ansatz.

\begin{remark}
Although this work focuses on the one-dimensional NLDE, the proposed Dirac-sVQA framework provides a natural prototype for higher-dimensional nonlinear Dirac equations. Such an extension would rely on the same basic ingredients: a tensor-product position register for the multidimensional spatial grid, QFTs applied along each coordinate direction for the Fourier pseudospectral linear step, direction-wise signed Fourier-momentum encodings, and momentum-dependent spinor rotations constructed according to the chosen higher-dimensional Dirac representation. For the nonlinear part, many NLDEs retain local, state-dependent interactions expressed through spinor bilinears or densities; these terms may be incorporated through generalized measurement channels and variational updates, extending the self- and cross-channel D-QNPU primitives used here. A complete higher-dimensional implementation requires model-specific choices of spinor representation, nonlinear measurement decomposition, ansatz design, and resource analysis. We leave these developments to future work.
\end{remark}

\subsection{Variational formulation and measurable cost function}\label{subsec:cost}

Building on the SFDP and D-QNPU modules introduced above, we now cast the nonlinear update as a measurable variational optimization problem. The SFDP supplies the intermediate state $|\widetilde{\bm{\psi}}(t)\rangle$ after one structured linear Dirac step, while the D-QNPU provides the measurable nonlinear contributions required to evaluate the Dirac interaction. The remaining task is therefore to determine the parameters of a trial state at time $t+\tau$ that best approximates the nonlinear update of $|\widetilde{\bm{\psi}}(t)\rangle$.

Variational quantum algorithms are hybrid quantum-classical methods in which a parameterized cost function is minimized by a classical optimizer. Let $U(\bm{\theta})$ denote the variational ansatz circuit, where $\bm{\theta}\in\mathbb{R}^{N_p}$ is the parameter vector and $N_p$ is the number of trainable parameters. At time $t$, the current approximation is represented by
\begin{equation*}
|\bm{\psi}(t)\rangle
=
U(\bm{\theta}_t)|0\rangle^{\otimes(n+1)}.
\end{equation*}
After the linear Dirac substep, the intermediate state becomes
\begin{equation*}
|\widetilde{\bm{\psi}}(t)\rangle
=
\widetilde{U}(\bm{\theta}_t)|0\rangle^{\otimes(n+1)}.
\end{equation*}
The nonlinear variational step then seeks a trial state at time $t+\tau$ that approximates the state obtained by applying the short-time nonlinear map to the intermediate state, namely
\begin{equation*}
|\bm{\psi}(t+\tau)\rangle
\approx
Q[\widetilde{\bm{\psi}}(t)]\,|\widetilde{\bm{\psi}}(t)\rangle.
\end{equation*}

To determine the optimal parameters, we introduce the raw cost function
\begin{equation}\label{eq:dirac-cost-raw}
C(\bm{\theta})
=
\left\|
|\bm{\psi}(\bm{\theta})\rangle
-
Q[\widetilde{\bm{\psi}}(t)]\,|\widetilde{\bm{\psi}}(t)\rangle
\right\|^2.
\end{equation}
Expanding the square yields
\begin{equation}\label{eq:dirac-cost-raw-Re}
C(\bm{\theta})
=
C_1
-
2\,\mathrm{Re}\,
\langle \bm{\psi}(\bm{\theta}) |
Q[\widetilde{\bm{\psi}}(t)]
|\widetilde{\bm{\psi}}(t)\rangle,
\end{equation}
where $C_1$ is independent of $\bm{\theta}$ and therefore does not affect the minimizer. Substituting \eqref{eq:dirac-Q} into \eqref{eq:dirac-cost-raw-Re}, expanding the overlap term, and discarding additive constants independent of $\bm{\theta}$, we obtain
\begin{equation*}
C(\bm{\theta})
=
C_2
-
2\,\mathrm{Re}\,
\langle \bm{\psi}(\bm{\theta})|\widetilde{\bm{\psi}}(t)\rangle
-
2\tau\,
\mathrm{Im}\,
\langle \bm{\psi}(\bm{\theta})|
G[\widetilde{\bm{\psi}}(t)]
|\widetilde{\bm{\psi}}(t)\rangle.
\end{equation*}
Using the D-QNPU decomposition \eqref{eq:dirac-nonlinear-decomp}, the implementable cost function can therefore be written as
\begin{equation}\label{eq:dirac-cost-measurable}
C_{\mathrm{impl}}(\bm{\theta}_{t+\tau})
=
-2E_{\mathrm{ov}}
-
2\tau\,\frac{\gamma}{h}
\Big[
(\lambda_2+\lambda_1)E_{\mathrm{self}}
+
(\lambda_2-\lambda_1)E_{\mathrm{cross}}
\Big],
\end{equation}
where 
$E_{\mathrm{ov}}
=
\mathrm{Re}\,
\langle \bm{\psi}(t+\tau)|\widetilde{\bm{\psi}}(t)\rangle$, $E_{\mathrm{self}}$,  and $E_{\mathrm{cross}}$ are the self- and
cross-channel quantities defined in \eqref{eq:dirac-nonlinear-decomp}.
They depend on the trial state
$|\bm{\psi}(t+\tau)\rangle=U(\bm{\theta}_{t+\tau})|0\rangle^{\otimes(n+1)}$
and on the fixed intermediate state
$|\widetilde{\bm{\psi}}(t)\rangle$.

Accordingly, each evaluation of the nonlinear variational objective can be
reduced to the estimation of three measurement-accessible quantities:
\begin{itemize}
\item the overlap term $E_{\mathrm{ov}}$, estimated by a Hadamard-test or
swap-test-type circuit;
\item the self-channel contribution $E_{\mathrm{self}}$, obtained from the
self-channel measurement primitive of the D-QNPU;
\item the cross-channel contribution $E_{\mathrm{cross}}$, obtained from the
cross-channel measurement primitive of the D-QNPU.
\end{itemize}
Thus, the nonlinear Euler correction is reformulated as the minimization of a
measurement accessible cost function over the parameter space of the ansatz
circuit. This construction provides the operational link between the
operator-splitting discretization of the NLDE and the hybrid quantum-classical optimization loop of Dirac-sVQA.

To minimize the implementable cost function in practice, we employ the limited-memory Broyden-Fletcher-Goldfarb-Shanno algorithm with bound constraints (L-BFGS-B) \cite{byrd1995limited}. Although the cost functional contains the overlap term together with the two Dirac-specific nonlinear contributions, it remains a smooth finite-dimensional optimization problem in the ansatz parameters in the idealized numerical setting considered here. It is therefore well suited to a quasi-Newton method. L-BFGS-B incorporates approximate curvature information from successive iterations without explicitly forming the full Hessian, which typically improves convergence relative to plain gradient descent while remaining computationally tractable under repeated time stepping. The bound-constrained formulation is also natural for the rotation angles appearing in the parameterized ansatz circuit.

\subsection{Complexity analysis}\label{subsec:complexity}

We analyze the cost of Dirac-sVQA over one variational time step. We use four
measures: the circuit depth $C_D$, the gate complexity $C_G$, the query
complexity $C_Q$, and the total time complexity $C_T$. Here $C_Q$
denotes the number of quantum circuit executions, or equivalently the number
of measurement shots. We adopt the full-accounting convention
$C_T = C_G C_Q$, so that every measurement includes the cost of preparing the required quantum
states. Single-qubit rotations and controlled rotations are counted as
elementary gates; finite-precision synthesis would only add logarithmic
factors.

\paragraph{Ansatz complexity}
The Dirac data register contains $n+1$ qubits: $n$ position qubits and one
spin qubit. For a $d$-layer hardware-efficient ansatz with $O(n)$ gates per
layer, the conservative depth and gate estimates are
$C_D^{\mathrm{ans},D}=O(nd)$, and $C_G^{\mathrm{ans},D}=O(nd)$.
Thus, compared with a scalar amplitude encoding, the additional spin qubit
only changes the constants in the ansatz cost.

\paragraph{SFDP linear substep}
The structured linear Dirac substep is implemented by
$$U_{\mathrm{SFDP}}
=
R_\beta
(\mathrm{QFT})^\dagger
U_{\mathrm{ph}}^{(D)}
(\mathrm{QFT}).
$$
The two QFT blocks acting on the $n$-qubit position register require
$O(n^2)$ gates and depth under the standard exact QFT decomposition. The
Dirac phase block $U_{\mathrm{ph}}^{(D)}$ consists of $O(n)$
momentum-controlled spin rotations, and $R_\beta$ contributes only constant
cost. Therefore, we know
$C_D^{\mathrm{lin},D}=O(n^2)$ and $C_G^{\mathrm{lin},D}=O(n^2)$.

\paragraph{State preparation for nonlinear measurements}
For one cost-function evaluation at a trial parameter $\bm{\theta}$, the
measurement circuits require the linearly propagated state
$|a_t\rangle
=
U_{\mathrm{SFDP}}U(\bm{\theta}_t)|0\rangle^{\otimes(n+1)}$
and the trial state
$|b_{\bm{\theta}}\rangle
=
U(\bm{\theta})|0\rangle^{\otimes(n+1)}$.

Some measurement primitives may also use the conjugated ansatz state
\begin{equation*}
|b_{\bm{\theta}}^\ast\rangle
=
U^\ast(\bm{\theta})|0\rangle^{\otimes(n+1)},
\end{equation*}
which has the same asymptotic preparation cost as $ |b_{\bm{\theta}}\rangle $.
Thus, we have
$C_D^{a,D}=O(n^2+nd)$, $C_G^{a,D}=O(n^2+nd)$ and
$C_D^{b,D}=O(nd)$, $C_G^{b,D}=O(nd)$.
The additional D-QNPU operations for the self- and cross-channel primitives
consist of a constant or at most linear number of spin and ancilla-controlled
operations, depending on the chosen decomposition. These costs are lower order
compared with the preparation of $ |a_t\rangle $. Hence, per circuit
execution, we obtain
$C_D^{\mathrm{ov},D}= O(n^2+nd)$,
$C_D^{\mathrm{self},D}= O(n^2+nd)$,
$C_D^{\mathrm{cross},D} = O(n^2+nd)$,
and
$C_G^{\mathrm{ov},D} = O(n^2+nd)$,
$ C_G^{\mathrm{self},D}= O(n^2+nd)$,
$ C_G^{\mathrm{cross},D}=O(n^2+nd).$
The cross channel differs from the self channel only by an additional spin
flip, which does not change the asymptotic scaling.

\paragraph{Cost evaluation and query complexity}
The measurable cost function is
\begin{equation*}
C_{\mathrm{impl}}(\bm{\theta})
=
-2E_{\mathrm{ov}}
-
2\tau\frac{\gamma}{h}
\Big[
(\lambda_2+\lambda_1)E_{\mathrm{self}}
+
(\lambda_2-\lambda_1)E_{\mathrm{cross}}
\Big].
\end{equation*}
Suppose that each of the three observables is estimated with additive
statistical error at most $\delta$. Then
\begin{equation*}
\bigl|
\widehat{C}_{\mathrm{impl}}-C_{\mathrm{impl}}
\bigr|
\le
c_{\mathrm{meas}}\delta,
\end{equation*}
where 
$c_{\mathrm{meas}} = 2+ 2\tau\frac{\gamma}{h} \Big(|\lambda_2+\lambda_1|
+|\lambda_2-\lambda_1|\Big)$. To achieve a target additive accuracy $\varepsilon$ for the full cost, it is
sufficient to take $\delta = O(\varepsilon/c_{\mathrm{meas}})$.
With standard Monte Carlo sampling and bounded-variance measurement outcomes,
the query complexity for cost function evaluation is 
$C_Q^{\mathrm{cost},D}
=
O\!\left(c_{\mathrm{meas}}^2\varepsilon^{-2}\right)$.

The factor $c_{\mathrm{meas}}$ is kept explicit because the prefactor
$\gamma/h$ amplifies the contribution of the nonlinear observables in the
cost function. Although the unscaled self- and cross-channel observables may
be small under amplitude encoding, estimating their scaled contribution to a
fixed absolute accuracy still requires the corresponding precision in the
unscaled quantities, unless additional variance-reduction or rescaled
measurement procedures are introduced, it yields that
$C_T^{\mathrm{cost},D}
= O\!\left((n^2+nd)c_{\mathrm{meas}}^2\varepsilon^{-2}\right)$.
If the refinement regime satisfies $\tau\gamma/h=O(1)$ and the nonlinear
coefficients are bounded, then $c_{\mathrm{meas}}=O(1)$, and this simplifies
to
$C_T^{\mathrm{cost},D}
=
O\!\left((n^2+nd)\varepsilon^{-2}\right)$.

\paragraph{Complexity per time step}
Let $N_{\mathrm{it}}$ denote the number of cost-function evaluations used by
the classical optimizer in one time step, including any additional evaluations
needed for gradient estimation if applicable. Then the total time complexity
of one Dirac-sVQA time step is
$
C_T^{\mathrm{step},D}
=
O\!\left(
N_{\mathrm{it}}(n^2+nd)c_{\mathrm{meas}}^2\varepsilon^{-2}
\right).
$
Under the bounded-prefactor regime $c_{\mathrm{meas}}=O(1)$, this reduces to
$
C_T^{\mathrm{step},D}
=
O\!\left(
N_{\mathrm{it}}(n^2+nd)\varepsilon^{-2}
\right).
$

The estimates derived above are summarized in
Table~\ref{tab:dirac-complexity}. The table reports the depth 
and gate complexity per-shot circuit, the required query complexity, and the total 
time complexity resulting under the full-accounting convention. It shows that the
dominant deterministic cost per shot comes from the preparation of the linearly
propagated state, which contains both the ansatz circuit and the SFDP linear
block. The intrinsic D-QNPU measurement layer contributes only a lower-order
overhead to the circuit size. Thus, the leading cost of one Dirac-sVQA time
step is determined by SFDP based state preparation together with repeated
measurement sampling.

\begin{table}[htbp]
\centering
\caption{Asymptotic complexities of the main modules in Dirac-sVQA under the
full-accounting convention. Here $n$ is the number of position qubits,
$d$ denotes the ansatz depth, $\varepsilon$ is the target accuracy for the
full cost function, and $N_{\mathrm{it}}$ is the number of optimizer
cost-function evaluations per time step.}
\label{tab:dirac-complexity}
\resizebox{1\textwidth}{!}{%
\begin{tabular}{lcccc}
\toprule
Operation & $C_D$  & $C_G$  & $C_Q$ & $C_T$ \\
\midrule
Ansatz circuit $U(\bm{\theta})$
& $O(nd)$
& $O(nd)$
& $1$
& $O(nd)$ \\

SFDP linear block $U_{\mathrm{SFDP}}$
& $O(n^2)$
& $O(n^2)$
& $1$
& $O(n^2)$ \\

Propagated state preparation $|a_t\rangle$
& $O(n^2+nd)$
& $O(n^2+nd)$
& $1$
& $O(n^2+nd)$ \\

Observable measurements for
$E_{\mathrm{ov}},E_{\mathrm{self}},E_{\mathrm{cross}}$
& $O(n^2+nd)$
& $O(n^2+nd)$
& $O(c_{\mathrm{meas}}^2\varepsilon^{-2})$
& $O((n^2+nd)c_{\mathrm{meas}}^2\varepsilon^{-2})$ \\

Single cost-function evaluation
& $O(n^2+nd)$
& $O(n^2+nd)$
& $O(c_{\mathrm{meas}}^2\varepsilon^{-2})$
& $O((n^2+nd)c_{\mathrm{meas}}^2\varepsilon^{-2})$ \\

One Dirac-sVQA time step
& ---
& ---
& $O(N_{\mathrm{it}}c_{\mathrm{meas}}^2\varepsilon^{-2})$
& $O(N_{\mathrm{it}}(n^2+nd)c_{\mathrm{meas}}^2\varepsilon^{-2})$ \\
\bottomrule
\end{tabular}}
\end{table}

As shown in Table~\ref{tab:dirac-complexity}, the nonlinear measurement
primitives do not change the leading asymptotic circuit scaling under
full accounting, since each observable evaluation requires the same propagated
state preparation. Under a purely modular accounting, the standalone nonlinear
processing layer could be assigned a lower cost, such as $O(nd)$, which is
consistent with some early VQA conventions. However, for the present
Dirac-sVQA implementation, the full-accounting estimate in
Table~\ref{tab:dirac-complexity} gives a more faithful description of the
actual circuit cost.

\section{Circuit implementation of the SFDP and D-QNPU modules}\label{sec:circuit}

In this section, we present the explicit gate-level realization of the two core modules of Dirac-sVQA introduced in Section~\ref{sec:Dirac-sVQA}: the SFDP for the linear Dirac subflow and the Dirac quantum nonlinear processing unit (D-QNPU) for the measurable nonlinear update. The SFDP implements the structured unitary propagation associated with the linear Dirac operator on the joint position-spin register, whereas the D-QNPU provides the self- and cross-channel observables required by the variational cost function.

\paragraph{SFDP linear block}
We first describe the circuit realization of the linear Dirac substep. In the
Fourier pseudospectral representation, the position register is transformed to
the Fourier basis, where the derivative operator is diagonal in the momentum
index. For each Fourier mode, the kinetic Dirac term then acts as a
momentum-dependent rotation on the spin qubit. After applying this
momentum-controlled spin rotation, the state is transformed back to the
position basis, and the mass term is implemented as a local spin rotation.

To make the register structure explicit, let
$\mathcal F_x = \mathrm{QFT}\otimes \mathbb{I}_s$
denote the QFT acting only on the $n$-qubit position register. We write the
single-spin mass rotation as
\begin{equation*}
r_\beta
=
\exp(-i\beta \tau)
=
\exp(-i\sigma_z\tau)
=
R_z(2\tau),
\end{equation*}
where $R_z(\theta)=\exp(-i\theta\sigma_z/2)$. Its lift to the full
position-spin register is
$\mathcal R_\beta = \mathbb{I}_x\otimes r_\beta$.
With this notation, the SFDP circuit is
\begin{equation*}
U_{\mathrm{SFDP}}(\tau)
=
\mathcal R_\beta\,
\mathcal F_x^\dagger\,
U_{\mathrm{kin}}^{(D)}\,
\mathcal F_x .
\end{equation*}
Therefore, if $U(\bm{\theta}_t)$ prepares the spinor state at time $t$, the
complete intermediate-state preparation map is
\begin{equation}\label{eq:dirac-Ulin-circuit}
\widetilde{U}(\bm{\theta}_t)
=
U_{\mathrm{SFDP}}(\tau)\,U(\bm{\theta}_t)
=
\mathcal R_\beta\,
\mathcal F_x^\dagger\,
U_{\mathrm{kin}}^{(D)}\,
\mathcal F_x\,
U(\bm{\theta}_t).
\end{equation}

For a fixed Fourier momentum mode $\mu_k$, the kinetic Dirac contribution is
given by
\begin{equation}\label{eq:dirac-uphd-mode}
r_\alpha(\mu_k)
=
\exp(-i\alpha \mu_k\tau)
=
\exp(-i\sigma_x\mu_k\tau)
=
R_x(2\mu_k\tau),
\end{equation}
Hence the momentum-controlled kinetic block on
the full position-spin register is
\begin{equation}\label{eq:dirac-uphd}
U_{\mathrm{kin}}^{(D)}
=
\sum_{k=0}^{2^n-1}
|k\rangle\langle k|
\otimes
r_\alpha(\mu_k)
=
\sum_{k=0}^{2^n-1}
|k\rangle\langle k|
\otimes
R_x(2\mu_k\tau).
\end{equation}

The signed Fourier momentum is encoded directly by the binary digits of the
Fourier index, that is
\begin{equation}\label{eq:dirac-signed-momentum}
\mu_k
=
\frac{2\pi}{L}
\left(
\sum_{\ell=0}^{n-2}2^\ell b_\ell
-
2^{n-1}b_{n-1}
\right),
\qquad
b_\ell\in\{0,1\}.
\end{equation}
Thus the total spin-rotation angle decomposes into contributions from the
individual binary digits. Since all rotations are about the same spin axis,
the resulting controlled rotations commute, and $U_{\mathrm{kin}}^{(D)}$ can
be implemented as
\begin{equation}\label{eq:dirac-controlled-rx}
U_{\mathrm{kin}}^{(D)}
=
\prod_{\ell=0}^{n-2}
\Lambda_\ell
\left(
R_x\!\left(\frac{4\pi\tau}{L}2^\ell\right)
\right)
\,
\Lambda_{n-1}
\left(
R_x\!\left(-\frac{4\pi\tau}{L}2^{n-1}\right)
\right),
\end{equation}
where $\Lambda_\ell(R_x(\theta))$ denotes an $R_x(\theta)$ rotation on the
spin qubit controlled by the $\ell$-th position qubit. The negative sign in
the most significant bit implements the signed Fourier momentum convention,
so no explicit FFT-shift or additional $X_{\mathrm{MSB}}$ centering gates are
required.

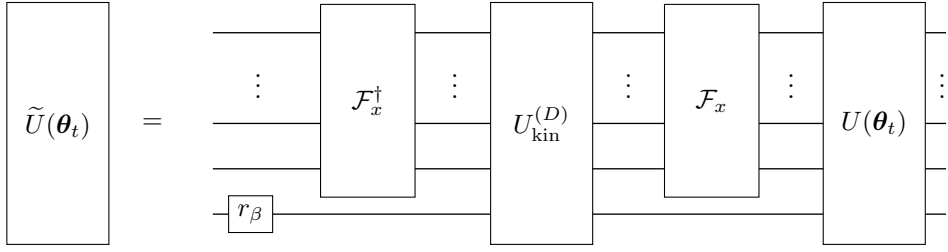
\begin{figure}[htbp]
\centering
\begin{tikzpicture}[x=1cm,y=1cm,>=latex]

\tikzstyle{posblock}=[draw, fill=white,minimum width=1.25cm,minimum height=2.55cm,align=center]
\tikzstyle{fullblock}=[draw, fill=white,minimum width=1.35cm,minimum height=3.20cm,align=center]
\tikzstyle{smallblock}=[draw, fill=white,minimum width=0.55cm,minimum height=0.45cm,align=center]
\tikzstyle{wire}=[line width=0.5pt]

\node[fullblock] (Uleft) at (0,0) {$\widetilde{U}(\bm{\theta}_t)$};
\node at (1.25,0) {$=$};

\def\yA{1.20}
\def\yB{0.60}
\def\yC{0.00}
\def\yD{-0.60}
\def\yS{-1.20}

\draw[wire] (2.05,\yA) -- (11.75,\yA);
\draw[wire] (2.05,\yC) -- (11.75,\yC);
\draw[wire] (2.05,\yD) -- (11.75,\yD);
\draw[wire] (2.05,\yS) -- (11.75,\yS);

\node[smallblock] (Rbeta) at (2.55,\yS) {$r_\beta$};
\node[posblock] (Rqftd) at (4.10,0.30) {$\mathcal F_x^\dagger$};
\node[fullblock] (UkinD) at (6.40,0) {$U_{\mathrm{kin}}^{(D)}$};
\node[posblock] (Rqft) at (8.65,0.30) {$\mathcal F_x$};
\node[fullblock] (Ulam) at (10.80,0) {$U(\bm{\theta}_t)$};

\node at (2.65,\yB) {$\vdots$};
\node at (5.25,\yB) {$\vdots$};
\node at (7.55,\yB) {$\vdots$};
\node at (9.70,\yB) {$\vdots$};
\node at (11.70,\yB) {$\vdots$};

\end{tikzpicture}
\caption{Overall circuit realization of
$\widetilde{U}(\bm{\theta}_t)=U_{\mathrm{SFDP}}(\tau)U(\bm{\theta}_t)$.
The circuit is read from right to left. The rightmost block prepares the input
spinor state, $\mathcal F_x=\mathrm{QFT}\otimes I_s$ maps the position
register to the Fourier basis, $U_{\mathrm{kin}}^{(D)}$ implements the
momentum-controlled kinetic spin rotation, and $\mathcal F_x^\dagger$
returns the state to the position basis. The leftmost gate
$r_\beta=R_z(2\tau)$ applies the mass-induced spin rotation on the spin
qubit; equivalently, its lift to the full position-spin register is
$\mathcal R_\beta=I_x\otimes r_\beta$.}
\label{fig:dirac-linear-graybox}
\end{figure}

The controlled-rotation structure realizes the momentum-dependent spin rotation
associated with the kinetic part of the linear Dirac operator at the gate level,
while \(\mathcal R_\beta\) implements the local spin-phase evolution generated
by the mass term.
Figure~\ref{fig:dirac-linear-graybox} shows the resulting circuit for the
intermediate map $\widetilde{U}(\bm{\theta}_t)$. Reading the circuit from
right to left, the ansatz $U(\bm{\theta}_t)$ prepares the input spinor state,
$\mathcal F_x$ maps the position register to the Fourier basis,
$U_{\mathrm{kin}}^{(D)}$ applies the momentum-controlled spin rotation,
$\mathcal F_x^\dagger$ returns the state to the position basis, and
$\mathcal R_\beta$ applies the mass rotation on the spin register.

\paragraph{D-QNPU nonlinear block}
We next describe the circuit realization of the measurable nonlinear contribution
appearing in the cost function \eqref{eq:dirac-cost-measurable}. As discussed
in Section~\ref{subsec:D-QNPU}, the state-dependent nonlinear map is not
implemented directly as a fixed unitary. Instead, the D-QNPU estimates the
self- and cross-channel observables that enter the variational update.

Let
\begin{equation*}
|a\rangle
:=
|\widetilde{\bm{\psi}}(t)\rangle
=
\sum_{j=0}^{M-1}\sum_{\sigma=0}^{1}
a_{j,\sigma}|j,\sigma\rangle,
\qquad
|b\rangle
:=
|\bm{\psi}(t+\tau)\rangle
=
\sum_{j=0}^{M-1}\sum_{\sigma=0}^{1}
b_{j,\sigma}|j,\sigma\rangle,
\end{equation*}
where $a_{j,\sigma}:=\widetilde{\psi}_{j,\sigma}(t)$,
and  $b_{j,\sigma}:=\psi_{j,\sigma}(t+\tau)$. We further introduce the amplitude-wise complex conjugate state
$|b^\ast\rangle
=
\sum_{j=0}^{M-1}\sum_{\sigma=0}^{1}
b_{j,\sigma}^{\ast}|j,\sigma\rangle$.

Since the state-dependent nonlinear operator
\(G[\widetilde{\bm{\psi}}(t)]\) is diagonal in the composite basis
\(\{|j,\sigma\rangle\}\), it holds that
$G[\widetilde{\bm{\psi}}(t)]|j,\sigma\rangle
=
g_{j,\sigma}|j,\sigma\rangle$, with
$g_{j,\sigma}
=
\frac{\gamma}{h}
\left(
\lambda_2 \rho_j+\lambda_1 s_j z_\sigma
\right)$.
Here $\rho_j=|a_{j,0}|^2+|a_{j,1}|^2$,
and $s_j=|a_{j,0}|^2-|a_{j,1}|^2$, and \(z_0=1\), \(z_1=-1\) are the eigenvalues of the Pauli-\(Z\) operator associated with the two spin components. Equivalently, for
\(\sigma\in\{0,1\}\), the diagonal coefficient can be decomposed as
\begin{equation*}
\lambda_2\rho_j+\lambda_1 s_j z_\sigma
=
(\lambda_2+\lambda_1)|a_{j,\sigma}|^2
+
(\lambda_2-\lambda_1)|a_{j,1-\sigma}|^2 .
\end{equation*}
Therefore, the nonlinear overlap can be written as
\begin{equation}
\label{eq:IMESELF}
\operatorname{Im}
\langle b|
G[\widetilde{\bm{\psi}}(t)]
|a\rangle
=
\frac{\gamma}{h}
\operatorname{Im}\!\left[
\sum_{j=0}^{M-1}\sum_{\sigma=0}^{1}
b_{j,\sigma}^{\ast}a_{j,\sigma}
\left(
(\lambda_2+\lambda_1)|a_{j,\sigma}|^2
+
(\lambda_2-\lambda_1)|a_{j,1-\sigma}|^2
\right)
\right].
\end{equation}

The QNPU is based on the circuits from \cite{jaksch2023variational,Lubasch2020Nonlinear} for evaluating functions $\mathscr{F}$ of the form
$\mathscr{F} = f^{(1)\ast}\prod_{\ell=1}^{r} f^{(\ell)} $. The product index \(\ell\) is used here to distinguish it from the spatial grid
index \(j\). In the present Dirac-spinor setting, the amplitude functions are
defined on the composite index \((j,\sigma)\). it yields that
$\mathscr{F}_{j,\sigma}
=
f^{(1)\ast}_{j,\sigma}
\prod_{\ell=1}^{r}
f^{(\ell)}_{j,\sigma}$.

With the appropriate phase gate on the ancilla, the D-QNPU circuit estimates the
imaginary part of the corresponding global sum $\operatorname{Im}
\sum_{j=0}^{M-1}\sum_{\sigma=0}^{1}
\mathscr{F}_{j,\sigma}$.
For the cubic nonlinear term considered here, we take \(r=3\), we obtain
\begin{equation*}
\mathscr{F}_{j,\sigma}
=
f^{(1)\ast}_{j,\sigma}
f^{(1)}_{j,\sigma}
f^{(2)}_{j,\sigma}
f^{(3)}_{j,\sigma}
=
|f^{(1)}_{j,\sigma}|^2
f^{(2)}_{j,\sigma}
f^{(3)}_{j,\sigma}.
\end{equation*}
Thus, the measured \(r=3\) D-QNPU primitive is
$\operatorname{Im}
\sum_{j=0}^{M-1}\sum_{\sigma=0}^{1}
|f^{(1)}_{j,\sigma}|^2
f^{(2)}_{j,\sigma}
f^{(3)}_{j,\sigma}$ in the Dirac-spinor case. 
The self-channel contribution is obtained by choosing
$f^{(1)}=a$, $f^{(2)}=a$, and $f^{(3)}=b^\ast$, where \(b^\ast\) denotes the amplitude-wise complex conjugate of \(b\). This gives
\begin{equation}
\label{eq:Eself}
E_{\mathrm{self}}
:=
\operatorname{Im}
\sum_{j=0}^{M-1}\sum_{\sigma=0}^{1}
|a_{j,\sigma}|^2
a_{j,\sigma}
b_{j,\sigma}^{\ast}.
\end{equation}

For the cross-channel contribution, we introduce the spin-flipped state
\begin{equation*}
|a'\rangle := X_S |a\rangle,
\, {\rm and} \quad
X_S= \mathbb{I}_x^{\otimes n}\otimes X_s ,
\end{equation*}
where \(X_s\) acts only on the spin register. Consequently,
\begin{equation*}
a'_{j,\sigma}=a_{j,1-\sigma},
\, {\rm and} \quad
|a'_{j,\sigma}|^2=|a_{j,1-\sigma}|^2 .
\end{equation*}
By choosing $f^{(1)}=a'$, $f^{(2)}=a$, and $f^{(3)}=b^\ast$, the same \(r=3\) D-QNPU primitive yields
\begin{equation}
\label{eq:Ecross}
E_{\mathrm{cross}}
:=
\operatorname{Im}
\sum_{j=0}^{M-1}\sum_{\sigma=0}^{1}
|a_{j,1-\sigma}|^2
a_{j,\sigma}
b_{j,\sigma}^{\ast}.
\end{equation}

Above all, it follows from \eqref{eq:IMESELF}, \eqref{eq:Eself}, and \eqref{eq:Ecross}
that
\begin{equation}
\label{eq:nonlinear-self-cross-decomp}
\operatorname{Im}
\langle b|
G[\widetilde{\bm{\psi}}(t)]
|a\rangle
=
\frac{\gamma}{h}
\left[
(\lambda_2+\lambda_1)E_{\mathrm{self}}
+
(\lambda_2-\lambda_1)E_{\mathrm{cross}}
\right].
\end{equation}

\begin{figure}[htbp]
\centering
\resizebox{0.86\linewidth}{!}{
\begin{quantikz}[row sep=0.35cm, column sep=0.35cm]
\lstick{$c:\ket{0}$}
& \gate{H}
& \gate{S^\dagger}
& \qw
& \ctrl{2}
& \ctrl{3}
& \ctrl{1}
& \gate{H}
& \meter{} \\
\lstick{$R_1:\ket{0}^{\otimes(n+1)}$}
& \qw
& \gate{\widetilde{U}(\bm{\theta}_{t})}
& \gate{X_S}
& \qw
& \qw
& \gate[wires=3]{\textsf{QNPU}}
& \qw
& \qw \\
\lstick{$R_2:\ket{0}^{\otimes(n+1)}$}
& \qw
& \qw
& \qw
& \gate{\widetilde{U}(\bm{\theta}_{t})}
& \qw
& \qw
& \qw
& \qw \\
\lstick{$R_3:\ket{0}^{\otimes(n+1)}$}
& \qw
& \qw
& \qw
& \qw
& \gate{U^{\ast}(\bm{\theta}_{t+\tau})}
& \qw
& \qw
& \qw
\end{quantikz}}
\caption{Measurement circuit for the cross-channel observable
\(E_{\mathrm{cross}}\). The spin-flip gate
\(X_S=\mathbb{I}_x^{\otimes n}\otimes X_s\) is applied unconditionally to the
first data register to prepare \(|a'\rangle=X_S|a\rangle\).}
\label{fig:qnpu-cross}
\end{figure}
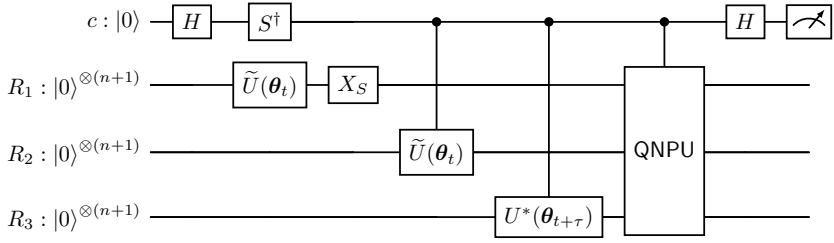

Operationally, the first D-QNPU data register supplies the density factor
\(|f^{(1)}_{j,\sigma}|^2\). Therefore, when evaluating the cross channel, the
spin-flipped state \(X_S|a\rangle\) must be prepared as the first input register.
The gate \(X_S\) is applied as an unconditional preprocessing gate to the first
data register, rather than as an ancilla-controlled operation. The corresponding
self- and cross-channel circuits are shown in Figs.~\ref{fig:qnpu-self} and
\ref{fig:qnpu-cross}, respectively.

\begin{figure}[htbp]
\centering
\resizebox{0.82\linewidth}{!}{
\begin{quantikz}[row sep=0.35cm, column sep=0.35cm]
\lstick{$c:\ket{0}$}
& \gate{H}
& \gate{S^\dagger}
& \ctrl{2}
& \ctrl{3}
& \ctrl{1}
& \gate{H}
& \meter{} \\
\lstick{$R_1:\ket{0}^{\otimes(n+1)}$}
& \qw
& \gate{\widetilde{U}(\bm{\theta}_{t})}
& \qw
& \qw
& \gate[wires=3]{\textsf{QNPU}}
& \qw
& \qw \\
\lstick{$R_2:\ket{0}^{\otimes(n+1)}$}
& \qw
& \qw
& \gate{\widetilde{U}(\bm{\theta}_{t})}
& \qw
& \qw
& \qw
& \qw \\
\lstick{$R_3:\ket{0}^{\otimes(n+1)}$}
& \qw
& \qw
& \qw
& \gate{U^{\ast}(\bm{\theta}_{t+\tau})}
& \qw
& \qw
& \qw
\end{quantikz}}
\caption{Measurement circuit for the self-channel observable
\(E_{\mathrm{self}}\). Each data wire denotes an \((n+1)\)-qubit register,
consisting of \(n\) position qubits and one spin qubit.}
\label{fig:qnpu-self}
\end{figure}
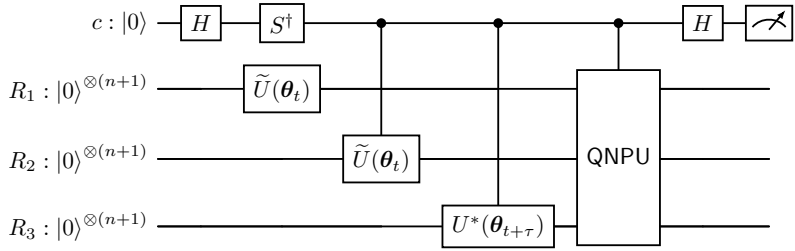

The D-QNPU measurement circuit uses three data registers \(R_1\), \(R_2\), and
\(R_3\), with one ancilla qubit \(c\). Each data register contains
\(n+1\) qubits, consisting of \(n\) position qubits and one spin qubit. The third
register is initialized in the amplitude-wise complex conjugate state
\(|b^\ast\rangle\). When the ansatz is composed of \(R_x\), \(R_z\), and CNOT
gates, the conjugate circuit \(U^\ast(\bm{\theta})\) is obtained by reversing
the signs of the rotation angles, while leaving the CNOT gates unchanged. This
makes the preparation of \(|b^\ast\rangle\) straightforward.

For the self channel, the three data registers are initialized as
$|a\rangle\otimes |a\rangle\otimes |b^\ast\rangle$.
Applying the \(r=3\) D-QNPU primitive and measuring the ancilla yields
\(E_{\mathrm{self}}\). For the cross channel, the first register is replaced by
the spin-flipped state \(|a'\rangle=X_S|a\rangle\), so that the three-register
input becomes
$|a'\rangle\otimes |a\rangle\otimes |b^\ast\rangle$.
Applying the same D-QNPU primitive then yields \(E_{\mathrm{cross}}\). Thus, the
only difference between the two nonlinear measurement circuits is the insertion
of the spin-flip gate \(X_S\) on the spin qubit of the first data register.
Together with the overlap observable
$E_{\mathrm{ov}}
:=
\operatorname{Re}
\big\langle
\bm{\psi}(t+\tau)
\big|
\widetilde{\bm{\psi}}(t)
\big\rangle$,
which can be estimated by a standard Hadamard test or swap test circuit,
the two D-QNPU outputs reproduce the implementable cost function introduced in
\eqref{eq:dirac-cost-measurable}, namely
\begin{equation*}
C_{\mathrm{impl}}(\bm{\theta}_{t+\tau})
=
-2E_{\mathrm{ov}}
-
2\frac{\tau \gamma}{h}
\left[
(\lambda_2+\lambda_1)E_{\mathrm{self}}
+
(\lambda_2-\lambda_1)E_{\mathrm{cross}}
\right].
\end{equation*}
Therefore, the nonlinear Dirac correction is reduced to three global expectation
values measured on structured circuits. In particular, the present construction
avoids sitewise reconstruction of local densities and spin polarizations, thereby
substantially reducing the measurement overhead and making the nonlinear step
compatible with the hybrid variational loop of Dirac-sVQA.

\section{Numerical experiments}\label{sec:numerics}

In this section, we investigate the numerical performance of the proposed Dirac-sVQA framework for the NLDE \eqref{eq:nlde}. The purpose of these experiments is threefold: first, to assess the accuracy of the SFDP and D-QNPU schemes in representative nonlinear regimes; second, to compare its behavior with corresponding classical splitting solvers; and third, to examine how the accuracy depends on the principal algorithmic parameters, namely the timestep size, the number of time steps, and the ansatz depth.

To probe the method under qualitatively distinct nonlinear interactions, we consider the following three representative coupling regimes:
\begin{itemize}
\item {Case I:} $\lambda_1=0$, $\lambda_2=10$;
\item {Case II:} $\lambda_1=10$, $\lambda_2=0$;
\item {Case III:} $\lambda_1=10$, $\lambda_2=10$.
\end{itemize}
These three cases isolate the density-dependent contribution, the spin-dependent contribution, and their combined action, respectively. They therefore provide a natural benchmark set for testing the nonlinear variational update implemented through the D-QNPU self- and cross-channel measurements.

To place the quantum results in context, we compare Dirac-sVQA with two classical Fourier-pseudospectral splitting solvers: \\
(i) the classical splitting (C) method defined by \eqref{eq:linear-step}--\eqref{eq:nonlinear-step}; \\
(ii) the same classical splitting method supplemented by an explicit normalization (NC) after each time step. \\
The comparison between C and NC isolates the role of norm control in the nonlinear regime, while the comparison between Dirac-sVQA and NC indicates how closely the variational quantum update reproduces the normalized splitting dynamics. In addition, the NC solver, when run at sufficiently fine resolution, is used to generate the reference solution for error evaluation.

We solve \eqref{eq:nlde} on the periodic domain $\bT=(-\pi,\pi)$, $\gamma=1$, and initial data
\begin{equation*}
\Psi^0(x)=\left(\frac{\sin(x)}{2+\cos^2(x)},\frac{\cos(x)}{2+\sin(x)}\right)^T, 
\qquad x\in \bT.
\end{equation*}
The spatial mesh size is $h=2\pi/M$ with $M=2^n$, where $n$ denotes the number of position qubits. In the quantum encoding, the position degree of freedom is stored on $n$ qubits, while one additional qubit encodes the two spinor components. Hence, each Dirac spinor state is represented on an $(n+1)$-qubit position-spin register. Unless stated otherwise, each Dirac-sVQA time step consists of an SFDP linear propagation followed by a D-QNPU-based variational nonlinear correction. All Dirac-sVQA numerical experiments are performed using Qiskit in the ideal
noiseless statevector-simulator setting. The cost function, including the
overlap term and the D-QNPU self- and cross-channel observables, is evaluated
from the full complex statevector, without finite-shot sampling, measurement
noise, or hardware noise. The classical variational optimization is carried out
with the Qiskit implementation of the L-BFGS-B algorithm, using at most $1200$
iterations and a stopping tolerance of $10^{-12}$.

Let $\Psi_{\mathrm{num}}(x_j,t_m)$ denote the numerical solution obtained with mesh size $h$ at point $x_j$ and timestep $\tau$ at time $t_m=m\tau$, $m=0,1,\dots$. To quantify the approximation error, we use the root mean square error
\begin{equation}\label{RMSE}
\mathrm{RMSE}(t_m)
=
\sqrt{
\frac{1}{M}\sum_{j=0}^{M-1}
\left(
|\Psi_{\mathrm{num}}(x_j,t_m)|
-
|\Psi_{\mathrm{ref}}(x_j,t_m)|
\right)^2},
\end{equation}
where $x_j=-\pi+jh$, and $|\Psi(x,t)|=\sqrt{|\Psi_0(x,t)|^2+|\Psi_1(x,t)|^2}$.
Here $\Psi_{\mathrm{ref}}$ is the reference solution generated by the NC method on a much finer grid and timestep, specifically with $h=2\pi/2^{10}$, and $\tau=0.01/2^7$.

\subsection{Accuracy and error measures}\label{subsec:accuracy-errors}

Because the NLDE \eqref{eq:nlde} is a two component spinor system rather than a scalar wave equation, the quality of the approximation cannot be judged adequately from the total density alone. We therefore monitor the componentwise densities $|\Psi_0(x,t)|^2$ and $|\Psi_1(x,t)|^2$ together with the total density
\begin{equation*}
\rho(x,t)=|\Psi_0(x,t)|^2+|\Psi_1(x,t)|^2.
\end{equation*}
These observables allow us to assess not only the transport of total probability density, but also the redistribution of mass between the two spinor components.

\begin{figure}[htbp]
\centering
    \includegraphics[width=0.49\textwidth]{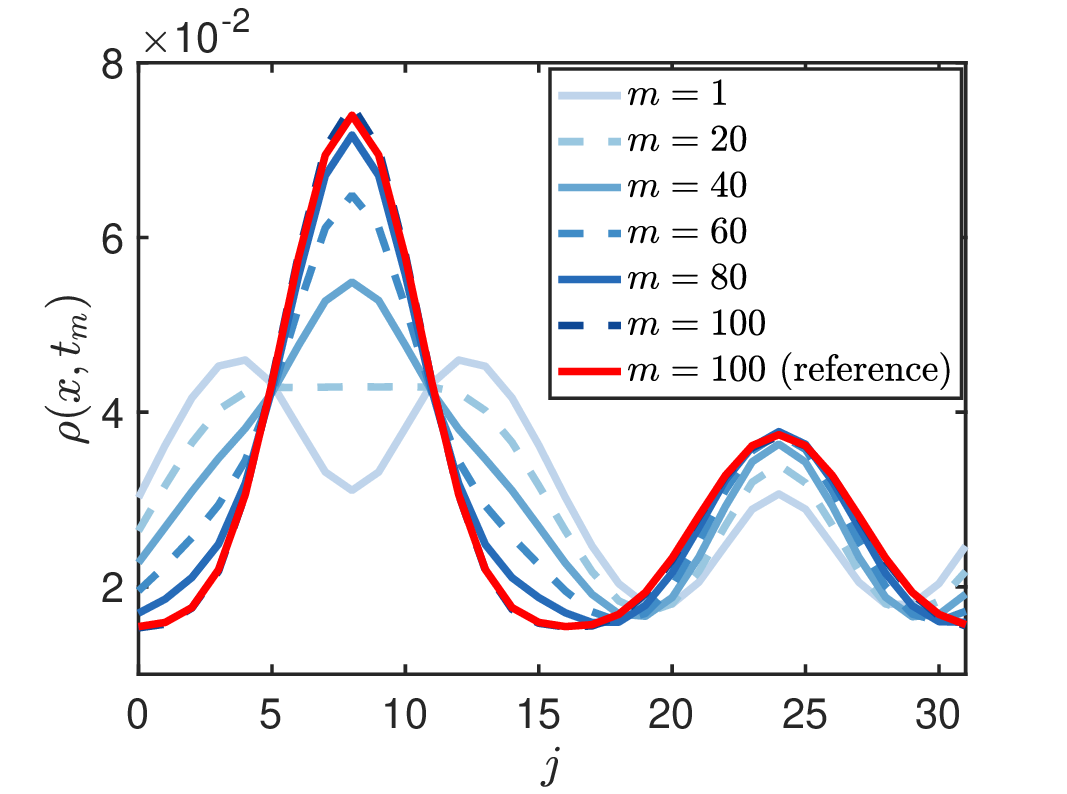}\\
    \includegraphics[width=0.49\textwidth]{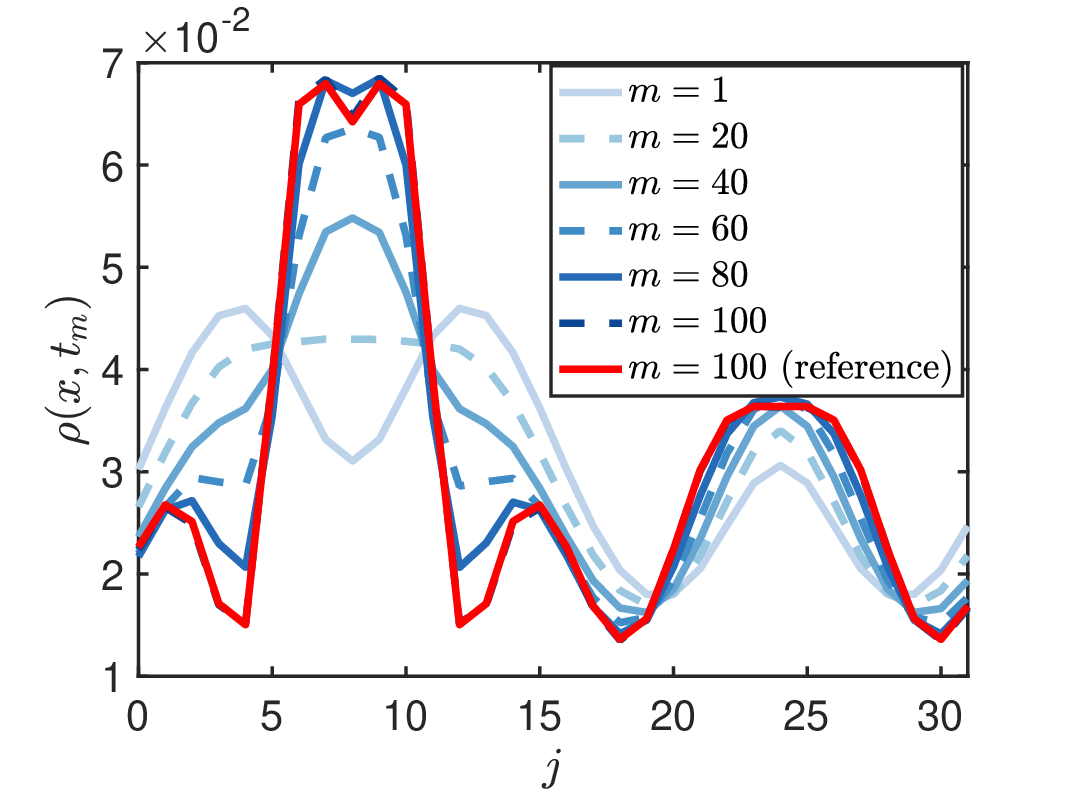}
    \includegraphics[width=0.49\textwidth]{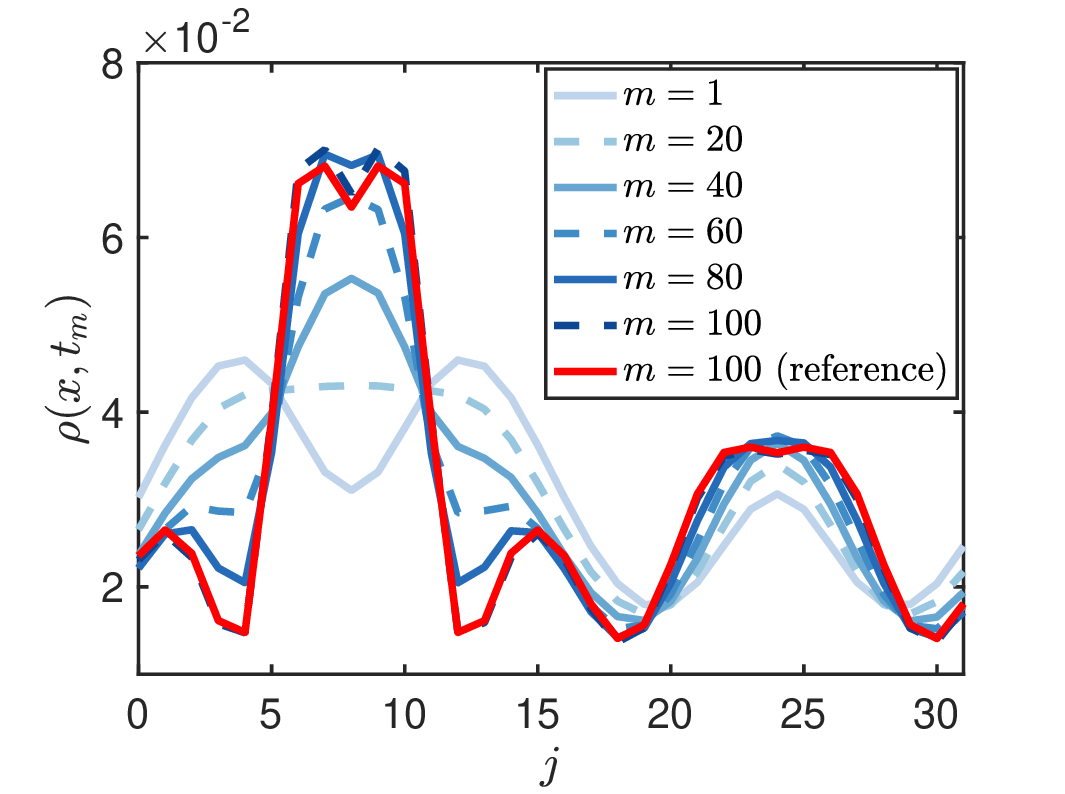}
\caption{Representative Dirac-sVQA solutions for the three benchmark cases. The three panels correspond to Case I, Case II, and Case III, respectively. For clarity, only representative final-time profiles are shown.}
\label{fig:dirac-wave-case8}
\end{figure}

We first examine the qualitative accuracy of the computed solution profiles. Figure~\ref{fig:dirac-wave-case8} shows representative Dirac-sVQA results for Cases I-III obtained with $n=5$, ansatz depth $d=8$, and timestep $\tau=0.005$. The numerical spinor profiles remain in very close agreement with the reference solution throughout the simulated interval. For visual clarity, only representative final-time profiles are displayed in the figure. These results show that the proposed SFDP and D-QNPU framework reproduces the main features of the nonlinear Dirac dynamics over a substantial propagation range.

\begin{figure}[htbp]
\centering
    \includegraphics[width=0.325\textwidth]{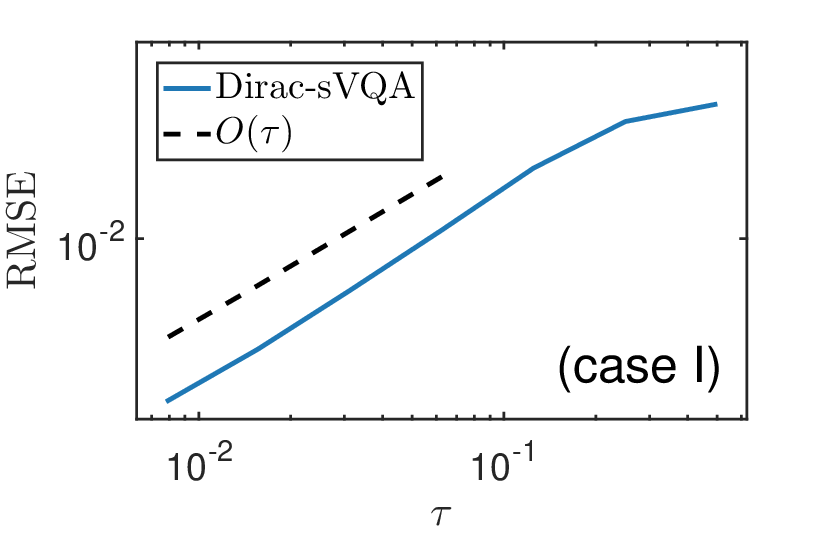}
    \includegraphics[width=0.325\textwidth]{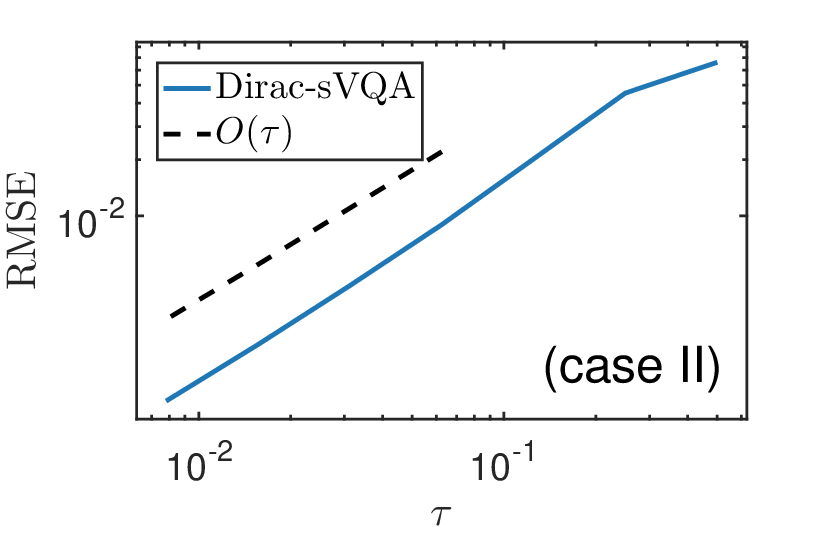}
    \includegraphics[width=0.325\textwidth]{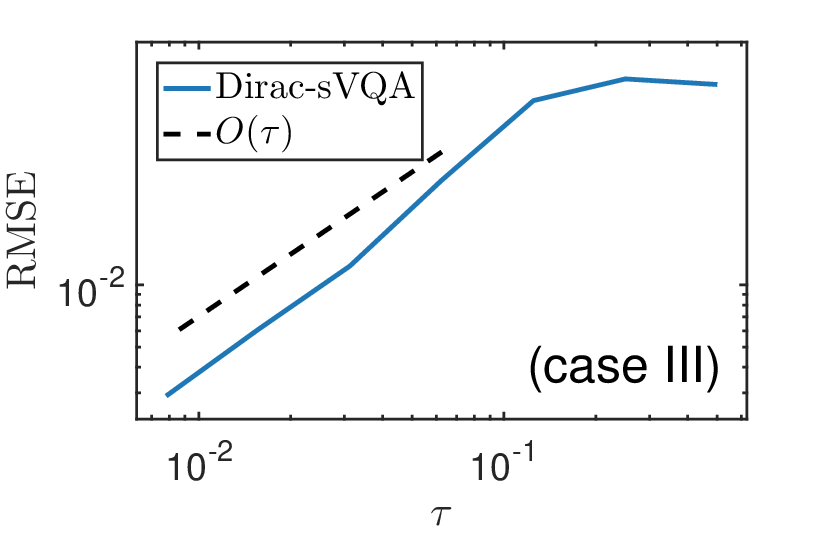}
\caption{Temporal accuracy of Dirac-sVQA for the three nonlinear benchmark cases. The first, second, and third columns correspond to Case I, Case II, and Case III, respectively. }
\label{fig:dirac-time-error}
\end{figure}

We next focus on the temporal accuracy of Dirac-sVQA. To make the temporal error dominant, we fix the spatial mesh at $h=2\pi/2^8$, so that the spatial discretization error is negligible compared with the time discretization error. Figure~\ref{fig:dirac-time-error} reports the RMSE at final time $t=0.5$ as a function of the timestep size for the three benchmark cases. In all three cases, the error exhibits approximately first-order convergence, which is consistent with the first-order nature of the split-step discretization combined with the explicit nonlinear Euler update.

\begin{figure}[htbp]
\centering
    \includegraphics[width=0.325\textwidth]{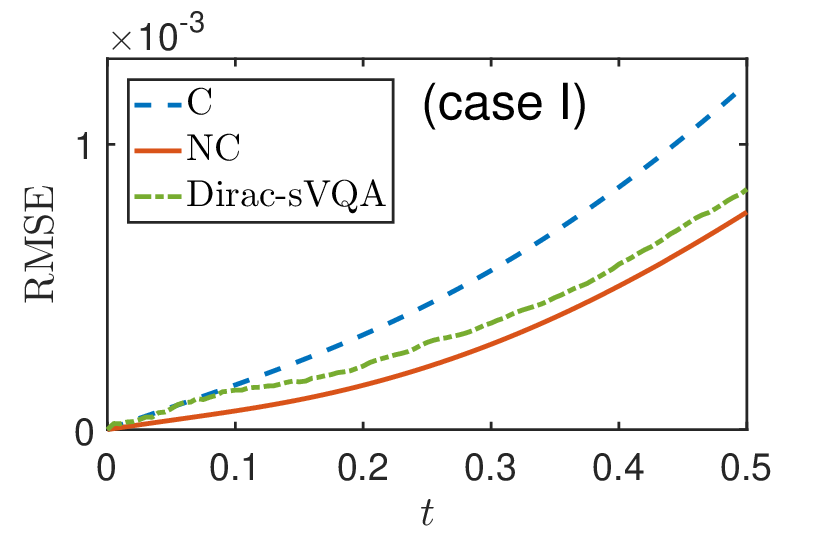}
    \includegraphics[width=0.325\textwidth]{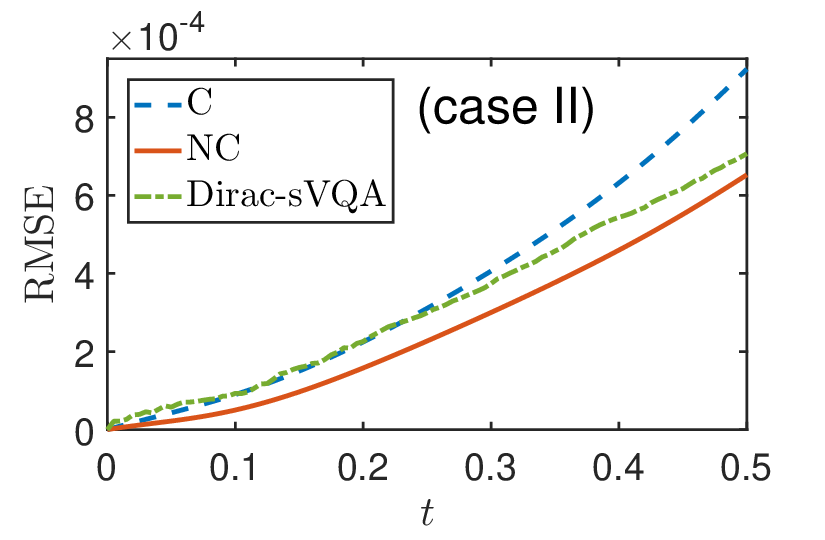}
    \includegraphics[width=0.325\textwidth]{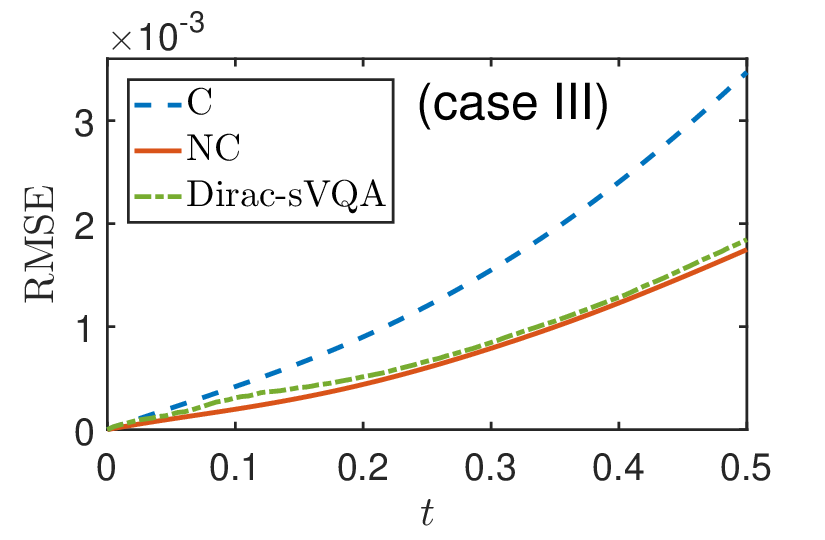}\\
\caption{Temporal evolution of the RMSE for the three benchmark cases. The first, second, and third columns correspond to Case I, Case II, and Case III, respectively. 
}
\label{fig:dirac-error}
\end{figure}

To compare the quantum and classical solvers more systematically, Fig.~\ref{fig:dirac-error} shows the temporal evolution of the RMSE for Dirac-sVQA, C, and NC using the same parameters $n=5$, $d=8$, and $\tau=0.005$. Across all three benchmark cases, the RMSE of Dirac-sVQA remains stable and well controlled over the entire simulation interval. Moreover, the Dirac-sVQA curves track the normalized classical scheme NC very closely, while remaining consistently more accurate than the unnormalized classical scheme C. This behavior is particularly significant in the fully coupled regime of Case III, where the effect of nonlinear interactions is strongest. Overall, these results indicate that the variational quantum update inherits the stabilizing effect of norm-preserving dynamics and reproduces the normalized splitting evolution much more faithfully than the raw explicit classical update.

\subsection{Expressibility of the Dirac ansatz}\label{subsec:expressibility}

We next examine whether the chosen parameterized quantum circuit has sufficient representational capacity for the spinor-valued states arising in the Dirac dynamics considered here. This question is especially relevant for the NLDE because the variational states are not scalar amplitudes, but two-component spinors encoded on a composite position-spin register. Accordingly, the ansatz must capture not only spatial amplitude variation, but also relative spinor phases, inter-component redistribution, and the correlations induced between spin and position degrees of freedom $N_{\mathrm{MC}}$.

To quantify expressibility, we adopt the standard fidelity-distribution approach and compare the distribution generated by the parameterized circuit with the Haar-random reference distribution on the same Hilbert space. Let $p_{\mathrm{PQC}}(F)$ denote the empirical fidelity distribution induced by the ansatz \cite{kocher2025numerical}, and let $p_{\mathrm{Haar}}(F)$ denote the corresponding Haar-random reference distribution \cite{Zyczkowski2005}. We then measure expressibility by the Kullback-Leibler divergence \cite{sim2019}
\begin{equation*}
D_{\mathrm{KL}}
=
\int_0^1
p_{\mathrm{PQC}}(F)
\log\!\left(
\frac{p_{\mathrm{PQC}}(F)}{p_{\mathrm{Haar}}(F)}
\right)\,dF.
\end{equation*}
Smaller values of $D_{\mathrm{KL}}$ indicate that the state distribution generated by the ansatz is closer to the Haar-random distribution and hence that the ansatz is more expressive in the global Hilbert-space sense.

\begin{figure}[htbp]
\centering
    \includegraphics[width=0.55\textwidth]{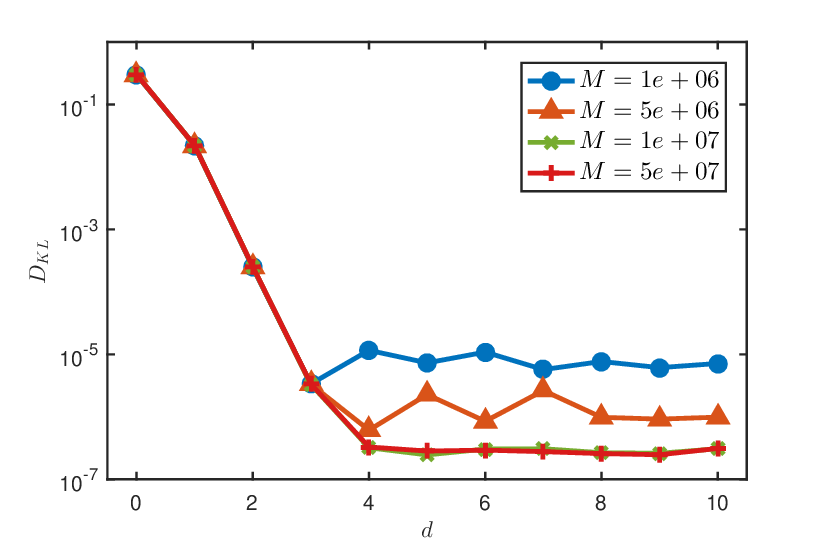}
\caption{Kullback-Leibler divergence $D_{\mathrm{KL}}$ as a function of the ansatz depth $d$ for several Monte Carlo sample sizes.}
\label{fig:dirac-expressibility}
\end{figure}

Figure~\ref{fig:dirac-expressibility} shows $D_{\mathrm{KL}}$ as a function of the circuit depth $d$ for several Monte Carlo sample sizes. The divergence decreases rapidly as the depth increases from very shallow values and then enters a clear saturation regime. This indicates that the chosen ansatz quickly acquires substantial global expressibility on the spin-position Hilbert space. In particular, for $n=5$, the Haar-based expressibility diagnostic becomes small already at relatively shallow depth.

At the same time, this expressibility test should be interpreted as an architectural diagnostic rather than as a direct predictor of the final task-specific numerical error. Indeed, the later accuracy study shows that the Dirac-sVQA error can continue to improve up to moderate depths such as $d\approx 8$, even though $D_{\mathrm{KL}}$ has already saturated much earlier. This is consistent with the well-known fact that Haar-based expressibility measures global coverage of Hilbert space, whereas actual variational performance also depends on the structure of the target states, the optimization landscape, and the accumulation of errors over repeated time stepping. In this sense, Fig.~\ref{fig:dirac-expressibility} supports the use of moderate circuit depths, while the final depth choice must ultimately be guided by task-oriented error measurements.

\subsection{Dependence on timestep size and ansatz depth}\label{subsec:parameter-dependence}

We now study how the numerical accuracy depends on the principal algorithmic parameters varied in the experiments: the timestep size $\tau$ and the ansatz depth $d$. For fixed final time $t$, varying $\tau$ probes the balance between temporal discretization error and the cumulative error introduced by repeated variational updates. Varying $d$, in turn, probes the representational power of the ansatz and helps identify the point beyond which additional circuit depth yields only marginal improvement.

\begin{figure}[htbp]
\centering
    \includegraphics[width=0.325\textwidth]{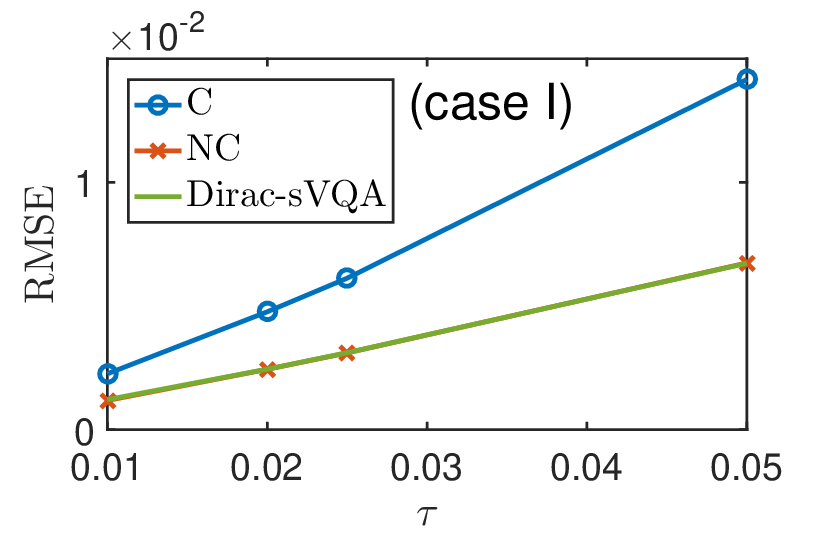}
    \includegraphics[width=0.325\textwidth]{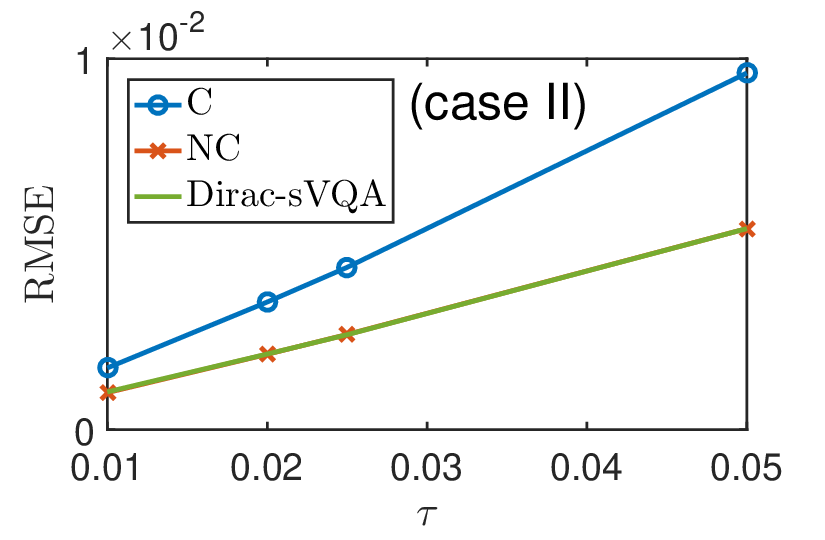}
    \includegraphics[width=0.325\textwidth]{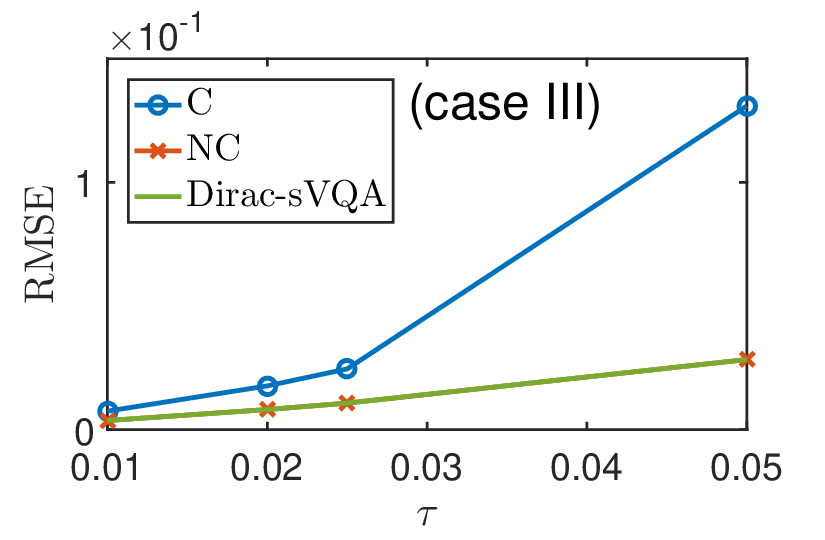}\\
\caption{RMSE as functions of the timestep size $\tau$ for the three benchmark cases with $d=8$. The first, second, and third columns correspond to Case I, Case II, and Case III, respectively.}
\label{fig:dirac-dt-case8}
\end{figure}

We begin with the dependence on $\tau$. Figure~\ref{fig:dirac-dt-case8} shows the RMSE as functions of the timestep size for the three benchmark cases. In all cases, the dependence on $\tau$ is clearly pronounced. Dirac-sVQA remains close to the normalized classical scheme NC over the entire parameter range, whereas the unnormalized classical scheme C deteriorates much more rapidly as $\tau$ increases. This confirms that norm control is crucial in the nonlinear regime and that the variational quantum formulation behaves much more like the normalized splitting solver than like the raw explicit update.

\begin{figure}[htbp]
\centering
    \includegraphics[width=0.325\textwidth]{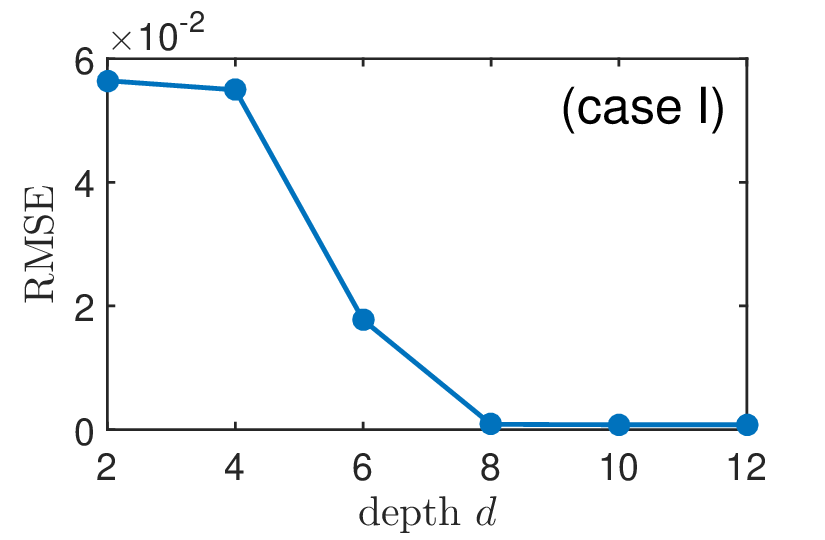}
    \includegraphics[width=0.325\textwidth]{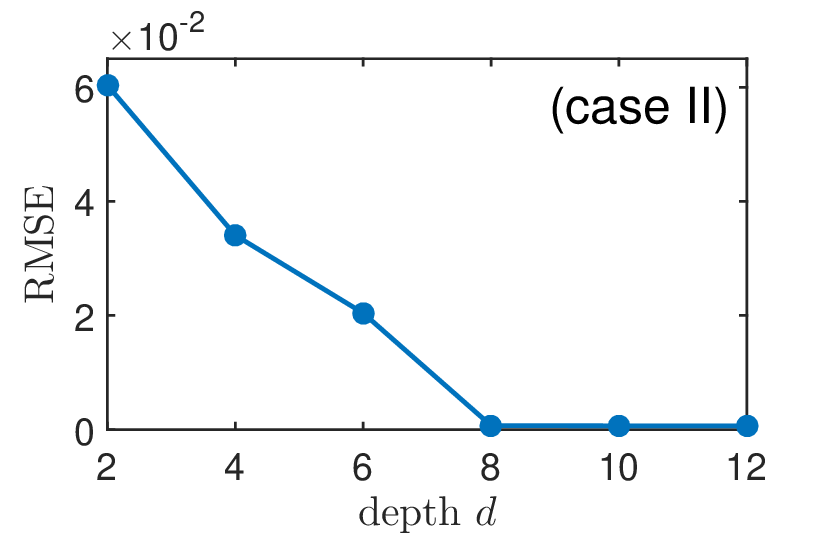}
    \includegraphics[width=0.325\textwidth]{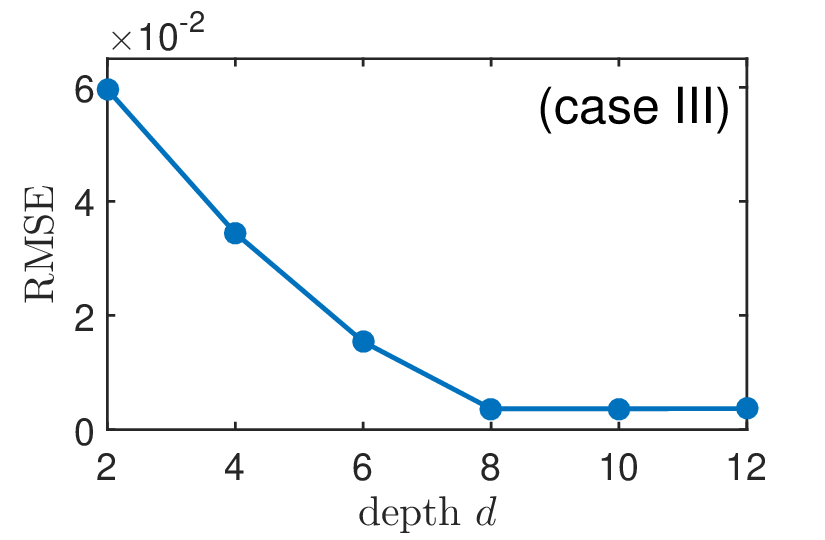}\\
\caption{RMSE as function of the ansatz depth $d$ for the three benchmark cases with $\tau=0.01$. The first, second, and third columns correspond to Case I, Case II, and Case III, respectively.}
\label{fig:dirac-depth-case8}
\end{figure}

We finally turn to the dependence on the ansatz depth. Here we fix $\tau=0.01$ and vary $d=2,4,6,8,10,12$. The results are shown in Fig.~\ref{fig:dirac-depth-case8}. Increasing the depth initially produces a substantial reduction in all error measures. The improvement is especially pronounced between $d=2$ and $d=8$, after which the curves largely level off. This indicates that moderate depths already provide sufficient representational power for the relevant nonlinear Dirac solution manifold, and that the remaining error is governed primarily by time discretization, repeated optimization, and measurement effects rather than by insufficient ansatz capacity alone. The observed saturation is also consistent with the expressibility study, although the depth at which the numerical error plateaus is naturally more task-dependent than the depth at which the global expressibility diagnostic saturates.

Overall, these experiments show that Dirac-sVQA resolves both the total density and the componentwise spinor profiles with controlled errors in all three benchmark regimes. These regimes correspond to the density-dependent, spin-dependent, and mixed nonlinear cases considered above. Over the simulated time interval, the error curves of Dirac-sVQA remain close to those of the normalized Fourier pseudospectral splitting solver, while the unnormalized explicit splitting scheme exhibits larger error growth. The timestep study confirms the expected sensitivity of the nonlinear dynamics to temporal resolution, and the depth study indicates that moderate circuit depths, such as $d=8$, are sufficient for the present tests, with only marginal improvement
observed at higher depths.


\section{Conclusion}\label{sec:conclusion}
We have proposed the Dirac-sVQA, an operator splitting variational quantum algorithm for the NLDE \eqref{eq:nlde}. The method combines a spinor-Fourier Dirac propagator (SFDP) for the linear Dirac substep with the Dirac quantum nonlinear processing unit (D-QNPU) for the state-dependent
nonlinear correction. The SFDP implements the momentum-dependent spin rotations and the mass-induced spin evolution, while the D-QNPU evaluates the nonlinear contribution through measurable overlap, self-spin, and cross-spin observables. This yields a hybrid quantum-classical formulation tailored to the spinor structure and matrix-valued operator structure of the NLDE \eqref{eq:nlde}. Numerical experiments indicate that Dirac-sVQA accurately captures both the total density and the componentwise spinor profiles in density-dependent, spin-dependent, and mixed nonlinear regimes. The results agree well with a normalized Fourier pseudospectral splitting solver and exhibit more stable error behavior than the unnormalized explicit splitting scheme. 
These findings suggest that splitting variational quantum methods provide a viable framework for nonlinear relativistic spinor dynamics. We will focus on rigorous error analysis, spinor-adapted ansatz design, noise-resilient and measurement-efficient implementations, and extensions to higher-dimensional systems in our further investigation.


\begin{thebibliography}{10}

\bibitem{Material}
{\sc D.~A. Abanin, S.~V. Morozov, L.~A. Ponomarenko, and et~al.}, {\em Giant nonlocality near the {D}irac point in graphene}, Science, 332 (2011), p.~328–330.

\bibitem{Alghassi2022FeynmanKac}
{\sc H.~Alghassi, A.~Deshmukh, N.~Ibrahim, N.~Robles, S.~Woerner, and C.~Zoufal}, {\em A variational quantum algorithm for the {F}eynman-{K}ac formula}, Quantum, 6 (2022), p.~730.

\bibitem{Ali2023PoissonVQA}
{\sc M.~Ali and M.~Kabel}, {\em Performance study of variational quantum algorithms for solving the {Poisson} equation on a quantum computer}, Phys. Rev. Applied, 20 (2023), p.~014054.

\bibitem{Alvarez1983}
{\sc A.~Alvarez and B.~Carreras}, {\em Solitons in the nonlinear {Dirac} equation}, Phys. Lett. A, 96 (1983), pp.~327--330.

\bibitem{Arrighi2013QWDirac}
{\sc P.~Arrighi, M.~Forets, and V.~Nesme}, {\em The {D}irac equation as a quantum walk: higher dimensions, observational convergence}, Journal of Physics A: Mathematical and Theoretical, 47 (2014), p.~465302.

\bibitem{nonre3}
{\sc W.~Bao, Y.~Cai, X.~Jia, and J.~Yin}, {\em Error estimates of numerical methods for the nonlinear {Dirac} equation in the nonrelativistic limit regime}, Sci. China Math., 59 (2016), pp.~1461--1494.

\bibitem{semire1}
{\sc J.~Bolte and K.~S}, {\em A semiclassical approach to the {Dirac} equation}, Ann. Phys., 274 (1999), pp.~125--162.

\bibitem{byrd1995limited}
{\sc R.~H. Byrd, P.~Lu, J.~Nocedal, and C.~Zhu}, {\em A limited memory algorithm for bound constrained optimization}, SIAM J. Sci. Comput., 16 (1995), pp.~1190--1208.

\bibitem{nldeua}
{\sc Y.~Cai and Y.~Wang}, {\em Uniformly accurate nested {Picard} iterative integrators for the {Dirac} equation in the nonrelativistic limit regime}, SIAM J. Numer. Anal., 57 (2019), pp.~1602--1624.

\bibitem{Cao2013Poisson}
{\sc Y.~Cao, A.~Papageorgiou, I.~Petras, J.~Traub, and S.~Kais}, {\em Quantum algorithm and circuit design solving the {Poisson} equation}, New J. Phys., 15 (2013), p.~013021.

\bibitem{Cerezo2021VQAReview}
{\sc M.~Cerezo, A.~Arrasmith, R.~Babbush, S.~Benjamin, S.~Endo, K.~Fujii, J.~McClean, K.~Mitarai, X.~Yuan, L.~Cincio, and P.~Coles}, {\em Variational quantum algorithms}, Nat. Rev. Phys., 3 (2021), pp.~625--644.

\bibitem{Childs2021HighPrecisionPDE}
{\sc A.~Childs, J.~Liu, and A.~Ostrander}, {\em High-precision quantum algorithms for partial differential equations}, Quantum, 5 (2021), p.~574.

\bibitem{Cuevas2014}
{\sc J.~Cuevas, P.~Kevrekidis, and F.~Williams}, {\em The nonlinear {Dirac} equation: a review of recent results}, Nonlinearity, 28 (2015), pp.~R123--R167.

\bibitem{DAriano2013QCADirac}
{\sc G.~M. D'Ariano, A.~Bisio, and P.~Perinotti}, {\em Derivation of the {D}irac equation from principles of information processing}, Phys. Rev. A., 90 (2014), p.~012106.

\bibitem{Dirac}
{\sc P.~Dirac}, {\em The quantum theory of the electron}, Proc. R. Soc. A, 117 (1928), p.~610–624.

\bibitem{DKG}
{\sc P.~D’Ancona, D.~Foschi, and S.~Selberg}, {\em Null structure and almost optimal local regularity for the {D}irac-{K}lein-{G}ordon system}, J. Eur. Math. Soc., 9 (2007), p.~877–899.

\bibitem{Fillion2017DiracQC}
{\sc F.~Fillion, S.~MacLean, and R.~Laflamme}, {\em Algorithm for the solution of the {Dirac} equation on digital quantum computers}, Phys. Rev. A, 95 (2017), p.~042343.

\bibitem{Gerritsma2010Dirac}
{\sc R.~Gerritsma, G.~Kirchmair, F.~Z{\"a}hringer, E.~Solano, R.~Blatt, and C.~Roos}, {\em Quantum simulation of the {Dirac} equation}, Nature, 463 (2010), pp.~68--71.

\bibitem{nonlinear12}
{\sc L.~Haddad and L.~Carr}, {\em The nonlinear {Dirac} equation in {Bose}-{Einstein} condensates: {Foundation} and symmetries}, Phys. D, 238 (2009), pp.~1413–--1421.

\bibitem{splitting1}
{\sc E.~Hairer, C.~Lubich, and G.~Wanner}, {\em {Geometric} {Numerical} {Integration}: {Structure-Preserving} {Algorithms} for {Ordinary} {Differential} {Equations}}, Springer, 2006.

\bibitem{HWYY2024}
{\sc Y.~He, Y.~Wang, J.~Z. Yang, and H.~Yin}, {\em Numerical methods for the nonlinear {Dirac} equation in the massless nonrelativistic regime}, East Asian J. Appl. Math., 14 (2024), pp.~79--103.

\bibitem{Hu2024QuantumCircuitsPDE}
{\sc J.~Hu, S.~Jin, N.~Liu, and L.~Zhang}, {\em Quantum circuits for partial differential equations via {Schr{\"o}dingerisation}}, Quantum, 8 (2024), p.~1563.

\bibitem{MDsplitting}
{\sc Z.~Huang, S.~Jin, P.~A. Markowich, C.~Sparber, , and C.~Zheng}, {\em A time-splitting spectral scheme for the {Maxwell-Dirac} system}, J. Comput. Phys., 208 (2005), pp.~761--789.

\bibitem{ingelmann2024two}
{\sc J.~Ingelmann, S.~Bharadwaj, P.~Pfeffer, K.~Sreenivasan, and J.~Schumacher}, {\em Two quantum algorithms for solving the one-dimensional advection-diffusion equation}, Comput. Fluids, 281 (2024), p.~106369.

\bibitem{jaksch2023variational}
{\sc D.~Jaksch, P.~Givi, A.~Daley, and T.~Rung}, {\em Variational quantum algorithms for computational fluid dynamics}, AIAA journal, 61 (2023), pp.~1885--1894.

\bibitem{Jin2024Schrodingerisation}
{\sc S.~Jin, N.~Liu, and Y.~Yu}, {\em Quantum simulation of partial differential equations via {Schr{\"o}dingerization}}, Phys. Rev. Lett., 133 (2024), p.~230602.

\bibitem{kocher2025numerical}
{\sc N.~K{\"o}cher, H.~Rose, S.~Bharadwaj, J.~Schumacher, and S.~Schumacher}, {\em Numerical solution of nonlinear {Schr{\"o}dinger} equation by a hybrid pseudospectral-variational quantum algorithm}, Sci. Rep., 15 (2025), p.~23478.

\bibitem{KSZ2021}
{\sc P.~Kr{\"a}mer, K.~Schratz, and X.~Zhao}, {\em Splitting methods for nonlinear {Dirac} equations with {Thirring}-type interaction in the nonrelativistic limit regime}, J. Comput. Appl. Math., 387 (2021), p.~112494.

\bibitem{Lubasch2020Nonlinear}
{\sc M.~Lubasch, J.~Joo, P.~Moinier, M.~Kiffner, and D.~Jaksch}, {\em Variational quantum algorithms for nonlinear problems}, Phys. Rev. A, 101 (2020), p.~010301.

\bibitem{MQ2002}
{\sc R.~Mclachlan and G.~Quispel}, {\em Splitting methods}, Acta Numer., 11 (2002), pp.~341--434.

\bibitem{Miyashita2023DiracQW}
{\sc S.~Miyashita, T.~Satoh, M.~Sugawara, N.~Benchasattabuse, K.~Nakanishi, M.~Hajdu{\v{s}}ek, H.~Choi, and R.~V. Meter}, {\em Digital quantum simulator for the time-dependent {D}irac equation using discrete-time quantum walks}, arXiv preprint arXiv:2305.19568,  (2023).

\bibitem{Montanaro2016}
{\sc A.~Montanaro}, {\em Quantum algorithms: An overview}, Quantum Inf., 2 (2016), p.~15023.

\bibitem{Sarma2024NonlinearMultiPDE}
{\sc A.~Sarma, T.~Watts, M.~Moosa, Y.~Liu, and P.~McMahon}, {\em Quantum variational solving of nonlinear and multidimensional partial differential equations}, Phys. Rev. A, 109 (2024), p.~062616.

\bibitem{nldelow}
{\sc K.~Schratz, Y.~Wang, and X.~Zhao}, {\em Low-regularity integrators for nonlinear {Dirac} equations}, Math. Comp., 90 (2021), pp.~189--214.

\bibitem{spectral}
{\sc J.~Shen, T.~Tang, and L.~Wang}, {\em Spectral Methods: Algorithms, Analysis and Applications}, Springer, 2011.

\bibitem{sim2019}
{\sc S.~Sim, P.~Johnson, and A.~Aspuru-Guzik}, {\em Expressibility and entangling capability of parametrized quantum circuits for hybrid quantum-classical algorithms}, Adv. Quantum Technol., 2 (2019), p.~1900070.

\bibitem{Tennie2025}
{\sc F.~Tennie, S.~Laizet, S.~Lloyd, and L.~Magri}, {\em Quantum computing for nonlinear differential equations and turbulence}, Nat. Rev. Phys., 7 (2025), pp.~1--11.

\bibitem{lambda}
{\sc W.~E. Thirring}, {\em A soluble relativistic field theory}, Ann. Physics., 3 (1958), pp.~91--112.

\bibitem{Xu2013NLDE}
{\sc J.~Xu, S.~Shao, and H.~Tang}, {\em Numerical methods for nonlinear {Dirac} equation}, J. Comput. Phys., 245 (2013), pp.~131--149.

\bibitem{Yuan2019}
{\sc X.~Yuan, S.~Endo, Q.~Zhao, Y.~Li, and S.~Benjamin}, {\em Theory of variational quantum simulation}, Quantum, 3 (2019), p.~191.

\bibitem{Zhang2025DiracSC}
{\sc Y.~Zhang, J.~Xu, Z.~Ma, W.~Zheng, D.~Lan, X.~Tan, and Y.~Yu}, {\em Experimental simulation of dirac equation in superconducting qubits}, Communications Physics,  (2025).

\bibitem{Zyczkowski2005}
{\sc K.~Zyczkowski and H.~Sommers}, {\em Average fidelity between random quantum states}, Phys. Rev. A, 71 (2005), p.~032313.

\end{thebibliography}
\end{document}